\documentclass[12pt,preprint]{aastex}
\usepackage{apjfonts}
\usepackage{epsfig,lscape}
\usepackage[dvipdfm]{hyperref}
\usepackage{natbib}

\newcommand{\hb}{\hbox{H$\beta$}}

\newcommand{\oii}{\hbox{[{O \sc ii}]}}
\newcommand{\nii}{\hbox{[{N \sc ii}]}}
\newcommand{\kms}{\hbox{km~s$^{-1}$}}

\newcommand{\nuvr}{\hbox{NUV$-r^{+}$}}
\newcommand{\nuvrcorr}{\hbox{$(\textmd{NUV}-r^{+})_{\rm corr}$}}
\newcommand{\oiii}{\hbox{[{O \sc iii}]}}

\shorttitle{GREEN GALAXIES IN COSMOS}
\shortauthors{Pan et al.}

\begin{document}

\title{GREEN GALAXIES IN THE COSMOS FIELD}

\author{Zhizheng Pan\altaffilmark{1,2}, Xu Kong\altaffilmark{1,2}
and Lulu Fan\altaffilmark{1,2}}
\email{panzz@mail.ustc.edu.cn, xkong@ustc.edu.cn}

\altaffiltext{1}{Center of Astrophysics, University of
Science and Technology of China, Hefei 230026, China}

\altaffiltext{2}{Key Laboratory for Research in Galaxies and Cosmology,
Chinese Academy of Sciences, Hefei 230026, China}


\begin{abstract}
We present a research of morphologies, spectra and environments of $\approx$
2350 ``green valley'' galaxies at $0.2<z<1.0$ in the COSMOS field. The
bimodality of dust-corrected \nuvr\ color is used to define ``green valley''
(thereafter, GV), which removes dusty star-forming galaxies from truly
transiting galaxies between blue cloud and red sequence. Morphological
parameters of green galaxies are intermediate between those of blue and red
galaxy populations, both on the Gini--Asymmetry and the Gini--M$_{\rm 20}$
planes. Approximately 60\% to 70\% green disk galaxies have intermediate or big
bulges, and only 5\% to 10\% are pure disk systems, based on the morphological
classification with Zurich Estimator of Structural Types (ZEST). The obtained
average spectra of green galaxies are intermediate between blue and red ones in
terms of \oii\,, H$\alpha$ and H$\beta$ emission lines. Stellar population
synthesis on the average spectra show that green galaxies are averagely older
than blue galaxies, but younger than red galaxies. Green galaxies have similar
projected galaxy density ($\Sigma_{10}$) distribution with blue galaxies at
$z>0.7$. At $z<0.7$, the fractions of $M_{\ast}<10^{10.0}M_{\sun}$ green
galaxies located in dense environment are found to be significantly larger than
those of blue galaxies. The morphological and spectral properties of green
galaxies are consistent with the transiting population between blue cloud and
red sequence. The possible mechanisms for quenching star formation activities
in green galaxies are discussed. The importance of AGN feedback cannot be well
constrained in our study. Finally, our findings suggest that environment
conditions, most likely starvation and harassment, significantly affect the
transformation of $M_{\ast}<10^{10.0}M_{\sun}$ blue galaxies into red galaxies,
especially at $z<0.5$.
\end{abstract}


\section{Introduction}

Many observational properties of galaxies in local universe, such as optical
colors \citep{Strateva 2001, Baldry 2004}, spectral indices \citep{Kauffmann
2003}, morphological parameters \citep{Driver 2006}, exhibit bimodal
distributions. Deeper surveys such as DEEP2 \citep{Willmer 2006}, COSMOS
\citep{Zamojski 2007} and NEWFIRM \citep{Brammer 2009} have revealed that the
bimodality of galaxy properties exists up to $z \approx 2$. Generally, galaxies
are categorized into two main populations: red quiescent galaxies and blue
star-forming galaxies. The red quiescent galaxies, mainly early-type galaxies
(ETGs), lie on a relative narrow ridge (the red sequence) in the
color-magnitude diagram (CMD). The blue star-forming galaxies, mainly late-type
galaxies (LTGs), are distributed throughout the so-called ``blue cloud".
Galaxies located in the region between blue cloud and red sequence are called
``green valley" galaxies (thereafter, green galaxies).

Studies on galaxy evolution reported that red galaxies have grown in stellar
mass by a factor of 2 to 4, whereas the number density of luminous blue
galaxies has dropped strongly since $z \approx$ 1 \citep{Bell 2004, Faber 2007,
Bolzonella 2010, Ilbert 2010, Brammer 2011}. The transformation of a
significant fraction of blue galaxies into red galaxies during this cosmic time
has been suggested by some previous works \citep{Faber 2007, Pozzetti 2010}.
However, the transiting timescale must be short \citep{Faber 2007, Martin 2007,
Balogh 2011}, since the bimodality will never be observed if the transiting
timescale is over a Hubble time. Green galaxies are therefore be considered as
a transiting population migrating towards the red sequence. However, only a few
studies have focused on this less prominent population. Some works reported
that a large fraction of green galaxies are not truly transiting, but are
dusty, star-forming galaxies \citep{Brammer 2009, Salim 2009}.

The physical properties of green galaxies have been revealed, to some extent,
in earlier works. Based on the DEEP2 redshift survey, \citet{Coil 2008} found
that the co-added spectra of green galaxies have higher \oiii/\hb\ ratio than
the spectra of blue galaxies at $z\approx 1$. Furthermore, the large-scale
clustering amplitude of green galaxies is similar to that of red galaxies.
However, green galaxies show similar clustering amplitude with blue galaxies at
a scale of $<1$ Mpc. Using SDSS main galaxy sample, \citet{Zehavi 2011} showed
that green galaxies have a clustering amplitude that can be found between
amplitudes of red and blue galaxies. \citet{Balogh 2011} reported that they
discovered a large transiting galaxy population in X-ray groups at $0.85<z<1.0$
in the COSMOS field, and these galaxies reside in the optical GV. Most of these
galaxies have bulges and small disk components, with half of them showing
prominent H$\delta$ absorption lines. \citet{Mendez 2011} studied the
morphologies of green galaxies in the AEGIS field and found that green galaxies
are intermediate between blue and red populations in terms of concentration,
asymmetry and morphological type; furthermore, removing dusty galaxies from the
green galaxy sample would not alter this result. \citet{Goncalves 2012} studied
the star formation history (SFH) of about 100 GV galaxies at $z \approx 0.8$
based on Keck spectra, and found that the transformation happened more rapidly
for more massive galaxies. Most previous studies are based on very small
samples, or focused on a very narrow redshift range. A specific study on
physical properties and redshift evolution of this population can be conducted
using a large and homogeneous sample that spans a wide redshift range, along
with deep multi-band photometry, spectra, high-resolution images, etc.

One of the emerged key questions is: which physical mechanisms are responsible
for quenching star formation activities in blue galaxies, then resulting the
transformation? Different star formation-quenching mechanisms have been
proposed. Of these, merger can effectively convert LTGs into ETGs, both in
observation and simulation. A major merger can change the overall galaxy
morphology as well as form bulge and trigger starburst, which can quickly
consume gas, or expel them through shocks from supernovae feedback
\citep{Springel 2005, Robertson 2006}. Another possible quenching mechanism is
the active galactic nucleus (AGN) feedback. In this scenario, the AGNs heat gas
to prevent them forming new stars \citep{Bower 2006, Tremonti 2007}. The AGN
feedback scenario is observationally supported by X-ray imaging of evacuated
cavities around massive galaxies \citep{McNamara 2000, McNamara 2007}.
Interestingly, some studies found that the AGN detection rate is high in GV
\citep{Nandra 2007, Coil 2009}. Meanwhile, environmental effects can lead to
quenching, as suggested in a great deal of works. For example, studies based on
{\it Hubble Space Telescope (HST)} showed that distant clusters have larger
fractions of LTGs than local clusters \citep{Dressler 1997, Fassano 2000, Lubin
2002}. These studies found the fraction of S0 galaxies has significantly
increased since $z\sim 0.7$, whereas the fraction of spiral galaxies decreased
significantly. The most favored explanation is that the dense environment can
effectively quench star formation activities in blue galaxies and turn them
into S0 galaxies \citep{Vogt 2004, Moran 2007}.

In this paper, we use the COSMOS data to study the transiting population and
constrain mechanisms that lead to star formation quenching. The COSMOS survey
\citep{Scoville 2007}, the largest contiguous survey with $HST$, provides a
combined data set including multi-band photometry covering X-ray, UV, optical,
infrared, millimeter to radio \citep{Bertoldi 2007,Capak 2007a,Hasinger
2007,Sanders 2007,Zamojski 2007}. The morphological parameters measured using
the Advanced Camera for Surveys (ACS) imaging are also available
\citep{Koekemoer 2007}. The zCOSMOS survey \citep{Lilly 2007} provides
approximately 10,000 galaxy spectra in 1.4 deg$^{2}$ COSMOS field, thereby
facilitating the exploration of spectral characters.The photometric redshifts
of COSMOS galaxies are extremely accurate \citep{Ilbert 2009}, and allows the
measurement of galaxy environment. Thus, the COSMOS provides an ideal data set
for selecting a large transiting galaxy sample, and for studying this sample in
detail.

This paper is organized as follows. Section 2 presents the data set used for
the study. Section 3 defines the red, green and blue galaxies, as well as the
mass limits. Section 4 presents the morphologies, stacked spectra and
environments of red, green and blue galaxies. Section 5 discusses the possible
mechanisms responsible for quenching star formation activities. A summary of
conclusions is presented in Section 6. Throughout this paper, a concordance
$\Lambda$CDM cosmology with $\Omega_{\rm m}=0.3$, $\Omega_{\rm \Lambda}=0.7$,
$H_{\rm 0}=70$ \kms Mpc$^{-1}$ is assumed. All magnitudes are quoted with the
AB normalization, unless explicitly noted.

\section{DATA}

\subsection{Photometric Redshift Catalog}

The large samples required for this work necessitate the use of photometric
redshifts. We use the COSMOS photometric redshift catalog (Version 1.8)
published by \citet{Ilbert 2009}. Accurate photometric redshifts were computed
by 30 broad, intermediate, and narrowband photometry, which included the UV
(\emph{Galaxy Evolution Explorer}), optical--NIR (Subaru, Canada France Hawaii
Telescope (CFHT), United Kingdom Infrared Telescope (UKIRT) and National
Optical Astronomy Observatory), and mid-IR (Spitzer/IRAC). \citet{Ilbert 2009}
computed the photometric redshifts of the COSMOS galaxies using \emph{Le
Phare}\footnote{www.oamp.fr/people/arnounts/LE\_PHARE.html} codes with a
$\chi^{2}$ template-fitting method, which included a novel treatment of the
emission lines. To generate the photometric catalog, all images had been
PSF-matched and fluxes were measured with SExtractor over an aperture with a
diameter of 3\arcsec\ at the position of Subaru $i^{+}$ band detection.
\citet{Ilbert 2009} included 9 templates generated by \citet{Polletta 2007} to
fit the UV-to-mid-IR data. To recover the SED of blue galaxies, 12 additional
templates were generated using BC03 model (Bruzual \& Charlot, 2003) with
starburst ages ranging from 0.03 to 3 Gyr. A comparison of the photometric and
spectroscopic redshifts showed that the accuracy is $\sigma_{\rm
(z_{phot}-z_{spec})}/(1+z_{\rm spec})$=0.009 for $i^{+}<22.5$ , $\sigma_{\rm
(z_{phot}-z_{spec})}/(1+z_{\rm spec})$=0.011 for $i^{+}<24.0$ and $\sigma_{\rm
(z_{phot}-z_{spec})}/(1+z_{\rm spec})$=0.057 for $24.0<i^{+}<25.0$ at $z<1.25$.
The SED template-fitting procedure also provided the best-fit rest-frame
absolute magnitudes and dust-corrected \nuvr\ color (thereafter \nuvrcorr).
Other galaxy physical properties, such as stellar mass and instantaneous
best-fit template star formation rate ($\rm SFR_{\rm template}$) were estimated
adopting BC03 models, with a Chabrier IMF and an exponentially declining star
formation history ($\rm SFR \propto e^{-t/\tau}$, $\tau$ in the
 range 0.1 -- 30 Gyr). To check the
robustness of $\rm SFR_{\rm template}$, \citet{Ilbert 2010} estimated the $\rm
SFR_{\rm IR}$ from IR luminosity $L_{\rm IR}$, where $L_{\rm IR}$ was
extrapolated from the Spitzer MIPS 24 $\mu$m flux using Dale \& Helow (2002)
library. $\rm SFR_{\rm template}$ and $\rm SFR_{\rm IR}$ were found to be
consistent upon comparison (see Figure 27 of Ilbert et al. 2010). In the
following sections, SFR refers to the best-fit model SFR ($\rm SFR_{\rm
template}$). More details please refer to \citet{Ilbert 2010}.

\subsection{Morphological catalog}
More than one million high-resolution galaxy images in COSMOS field are readily
available for the morphological studies, as observed with $HST$. In previous
works, automatic morphological classifications based on the parameters measured
on the initial galaxy images were used to improve efficiency and avoid
subjectivity \citep{Capak 2007b, Scarlata 2007, Tasca 2009}. Here we briefly
summarize the Zurich Estimator of Structure Types (ZEST) catalog \citep
{Scarlata 2007} used in this paper.

ZEST classifies galaxy morphology based on a principal component analysis (PCA)
of the galaxy structure nonparametric diagnostics, which include asymmetry $A$,
concentration $C$, Gini coefficient $G$, the second-order moment of the
brightest 20\% galaxy pixels \textmd{$ M_{\rm 20}$}, and the ellipticity of the
light distribution $\epsilon$. PCA shows that the first 3 PC variables
contribute more than 90\% of the original data set, thus can completely
describe galaxy structure. ZEST classifies galaxy morphologies into three main
types: elliptical galaxies ($T_{\rm ZEST}$=1), disk galaxies ($ T_{\rm
ZEST}$=2), and irregular galaxies ($T_{\rm ZEST}$=3). Disk galaxies are
assigned in 4 bins based on their ``bulgeness" parameter ($B_{\rm ZEST}$),
which roughly correlates with the bulge-to-disk ratio $[B/D]$. For galaxies
brighter than $i^{+}=22.5$, ZEST fits their surface brightness profile with
S\'{e}rsic model. Higher values of $ B_{\rm ZEST}$ indicate smaller bulges and
larger disks, with $ B_{\rm ZEST}$=[0,1,2,3], which correspond to the following
range of S\'{e}rsic index n: $n \geq 2.5$, $1.25\leq n \leq 2.5$, $0.75\leq n
\leq 1.25$ and $0\leq n \leq 0.75$, respectively. The final ZEST catalog
contains 131,532 sources down to $i^{+}=24.0$

The ZEST morphologies were tested using visual classification on $z=0$ sample.
It is found that the ZEST classifications agree well with the published
morphologies. Among disk galaxies with $ B_{\rm ZEST}$=0, approximately 85\%
are classified as elliptical and S0--Sab type.  Among disk galaxies with $
B_{\rm ZEST}$=2, the S0-Sab, Sb-Scd, and Sd types comprise 5\%, 45\% and 50\%,
respectively.The ZEST classifications begin to disintegrate at faint magnitude.
\citet{Scarlata 2007} quantified the efficiency of ZEST classifications by
degrading the signal-to-noise ratio of bright galaxies with well determined
morphologies. They reported that down to $i_{\rm 814w}$=22.5, the determination
of the final $T_{\rm ZEST}$ or $ B_{\rm ZEST}$ in 90\% galaxies does not
change. The fraction identified as the original morphological type is about
65\% down to $i_{\rm 814w}$=24.0.

\subsection{The 10k zCOSMOS Spectroscopic Sample}

The zCOSMOS spectroscopic survey \citep{Lilly 2007} targets the galaxies in
COSMOS field with the VIsible Multi-Object Spectrograph (VIMOS) mounted on the
ESO Very Large Telescope (VLT). The zCOSMOS survey has released about 10,000
galaxy spectra limited to $i^{+}=22.5$, covering a 1.4 deg$^{2}$ field of view.
The spectra range from 5500${\textmd{\AA}}$ to 9700${\textmd{\AA}}$, with a
medium resolution of $R \sim600$. Redshifts were measured by the VIPGI software
package. Results show that more than 97 \% spectra at $0.5<z<1.25$ can yield
very secure redshifts. Further information on zCOSMOS and the 10k sample are
presented in \citet{Lilly 2007}.

\section{Sample Selection}

\subsection{Green Galaxy Selection}

Our parent sample contains 113,162 galaxies extracted from the COSMOS
photometric redshift catalog (Version 1.8), with magnitudes and redshifts
limited to $i^{+}=24.0$ and $z_{\rm phot}=1.2$, respectively. The magnitude and
redshift limits ensure that our sample maintains high photometric redshift
accuracy. There are many methods to define GV, usually exploiting the natural
bimodal distribution of color indices such as $U-B$ \citep{Nandra 2007, Yan
2011}, $U-V$ \citep{Brammer 2009, Moresco 2010} and NUV$-r$ \citep{Wyder 2007}.
In many dusty starburst galaxies, large amount of internal dust absorbs
ultraviolet (UV) emission, resulting in red galaxy color. As mentioned in the
first section, a significant fraction of green galaxies are virtually
dust-obscured. To remove dusty star-forming galaxies from the real transiting
galaxy sample, we use dust-corrected \nuvr\ color to classify our sample.

Considering that considerable changes in extinction curves are expected from
galaxy to galaxy, to get suitable dust extinction, \citet{Ilbert 2009} used
different extinction curves depending on the best-fit template obtained in the
template-fitting procedure. \citet{Ilbert 2009} applied a test on the zCOSMOS
sample which with spectroscopic redshifts. They determined the best
fit-template color excess $E(B-V)^{\rm best}$ in the range of $E(B-V)^{\rm
best}=(0.0,0.5)$, using only passbands with $\lambda>3000(1+z)$\AA. As showed
in \citet{Ilbert 2009}, the $E(B-V)^{\rm best}$ value does not significantly
depend on the adopted extinction law. Then, they compared the rest-frame
observed magnitude $m^{\rm obs}$ and the predicted intrinsic magnitude without
extinction $m_{\rm uncor}^{\rm template}$ at $\lambda_{\rm
rest-frame}<3000$\AA, to discriminate between different extinction cures.
Figure.4 in \citet{Ilbert 2009} showed that for galaxies with template redder
than starburst template SB3, the $m^{\rm obs}-m_{\rm uncor}^{\rm
template}/E(B-V)^{\rm best}$ mainly follow the \citet{Prevot 1984} extinction
curve. For galaxies bluer than SB3, the \citet{Calzetti 2000} extinction law
was found to be more appropriate. \citet{Ilbert 2009} allowed an additional
2175\AA\ bump for the Calzetti law if a smaller $\chi^{2}$ is produced. No
reddening was allowed for galaxies redder than $Sb$.

Figure 1 shows \nuvrcorr\ as a function of redshift for
$M_{\ast}>10^{8.0}M_{\sun}$ galaxies. The most remarkable feature is that
\nuvrcorr\ clearly shows bimodal distribution up to $z\approx$ 1. In addition,
the global galaxy color shows significant evolution since $z\approx 1$, for
both red sequence galaxies and blue star-forming galaxies. Galaxies turn red
towards low redshifts, which demonstrates that the global star formation
activities decline with cosmic time \citep{Bell 2004, Brammer 2011}.

As the first step, we need to determine a criterion which can select
homogeneous green galaxies without bias to red or blue galaxies at all
redshifts. \citet{Ilbert 2010} classified galaxies into ``quiescent'' with
$\nuvrcorr\
>3.5$, ``high-activity" with $\nuvrcorr\ <1.2$ and ``intermediate activity''
with $1.2<\nuvrcorr\ <3.5$. As global galaxy color evolves significantly with
redshift, the sample selected with constant color cut over a wide redshift
range can be biased toward red and blue galaxies at high and low redshifts,
respectively. To minimize the bias, we have included the redshift evolution in
our green galaxies selection scheme. In Figure 1, two green lines represent our
selection criterion for green galaxy sample given by:
\begin{equation}
2.8-0.54z< \nuvrcorr\ <3.4-0.54z,
\end{equation}
where $z$ is the photometric redshift. Red and blue galaxies are defined with
$\nuvrcorr\ \geq 3.4-0.54z$ and $\nuvrcorr\ \leq 2.8-0.54z$, respectively. We
use a non-zero slope in our selection criterion. The slope $\alpha=-0.54$ came
from the linear fitting of median color for ``quiescent'' galaxies at 10
redshift bins within the range of $z= [0.2, 1.2] $. Our definition of green
galaxies is much stricter than the ``intermediate activity'' in \citet{Ilbert
2010} and considers the redshift evolution of galaxy color, which can help us
to compare the truly transiting population with active and quiescent
populations at the same cosmic epoch. The \nuvrcorr\ color has been found to be
well correlated with the specific star formation rate (sSFR) (defined as
$\rm{sSFR} = \rm{SFR}/M_{\ast}$) \citep{Salim 2005}. The median sSFR of our
green galaxy sample is about log(sSFR)$=-10.2$ at $z=1.0$ and log(sSFR)$=-10.8$
at $z=0.2$. We find that more than 95 \% galaxies in our sample lie on a very
narrow ridge in \nuvrcorr\ vs.  sSFR diagram. About 3\% of the galaxies have
high sSFR and red \nuvrcorr\,, indicating that these may be extremely dusty
starburst galaxies with underestimated dust extinction. We reject these
galaxies from further analysis. Likewise, the $z<0.2$ or $z>1.0$ green galaxies
were excluded because of the small sample size within these two redshift
ranges.

It is worth noting that at $z\sim 0.1$, the GV is centered at \nuvrcorr $\sim$
3.0, which seems to be much bluer than when it was firstly reported in several
GALEX papers \citep{Wyder 2007, Martin 2007}. \citet{Wyder 2007} studied the
UV-optical CMD using the GALEX and SDSS data, and found that GV is centered at
\nuvrcorr $\sim$ 4 (see Fig. 23 in Wyder et al 2007). The diversity is due to
the different data depth used in these two papers. The data in Wyder et al.
(2007) are limited to $r=17.6$, which is much shallower than that used in this
paper ($R_{\rm limit}\simeq 25.0$). Our sample is volume-completed for the
faintest source detected in Wyder et al. 2007 (about $M_{r}=-16.0$) out to $z
\sim 0.25$. Thus, the volume-corrected CMD in Wyder et al. (2007) should be
considered when comparing these two works. Keeping this in mind, excellent
consistency between Figure 1 in this study and the Figure 25 in Wyder et al.
(2007) is observed, which shows that the GV regions lie around \nuvrcorr $\sim
3$.

In a magnitude-limited sample, the minimum stellar mass for which observations
were completed depends on the redshift and stellar mass-to-light ratio ($M/L$).
Here, the low stellar mass limits were defined to reduce the fraction of
optical faint sources. To estimate the completeness mass we consider the
galaxies in the faintest 20\% of our sample and derive the stellar mass
($M_{\rm lim}$) they would have if their apparent magnitude were equal to the
limiting magnitude (i.e. $i^{+}=24.0$). Then we define as completeness mass the
value of the 95\% of the distribution of ($M_{\rm lim}$): galaxies above this
stellar mass limit define an 80\% complete sample in stellar mass (see Pozzetti
et al. 2010). We calculate the low mass limit for each redshift bin within the
redshift range of $z= [0.2, 1.0]$, with bin size of $\Delta z=0.1$. The values
of low mass limits for the COSMOS sample used in the study are similar with
those in literature \citep{Tasca 2009, Pozzetti 2010,Giodini 2012}.

Figure 2 shows the color-mass relation for galaxies within $z= [0.2, 1.0]$. To
separate the ``red sequence" and ``blue cloud" more clearly, we plot the
contours of galaxy number density. The vertical lines show the low mass limits,
which remove a larger fraction of active galaxies from the initial sample
compared with red and green galaxies. In the literature on stellar mass
functions, authors normally utilized different low mass limits for star-forming
and quiescent populations \citep{Pozzetti 2010, Ilbert 2010, Giodini 2012}. The
current paper focused on GV galaxies located in a relatively narrow region in
the CMD. For the green populations, their low mass limits do not significantly
depend on color. For blue galaxies, the low mass limits maitain 47.6\% and
12.8\% of the initial sample at $z= [0.2, 0.3]$ and $z= [0.9, 1.0]$,
respectively. For green galaxies, the mass limits maintain approximately 70\%
to 85\% of the initial sample in all redshift bins. The conclusion of the
present work is derived from the majority of green galaxies and are not
significantly affected by the low mass limits.

Figure. 2 shows that there is a remarkable deficit of
$M_{\ast}<10^{10.0}M_{\sun}$ red galaxies at z>0.6, which leads to difficulties
in defining the GV utilizing color-magnitude (or color-mass) relation, as shown
in some low-redshift studies.  Green galaxies are mainly distributed at the
minimum density regions in the diagram, or distribute at the reddest end of
``blue cloud", but these galaxies are till significantly redder than the bulks
of ``active" galaxies.

Figure 3 shows the color--color diagram ($M{\rm (NUV)}-M{\rm (r^{+})}$ versus
$M{\rm (r^{+})}-M{\rm (J)}$). Optical--near-IR color is prevalently used to
separate truly passive population from dusty star-forming galaxies
\citep{Williams 2009, Bundy 2010}. In Figure 3, red galaxies are located in a
tight clump in each redshift bin, which separate clearly from blue star-forming
galaxies. Green galaxies mainly reside in the joint regions between red and
blue galaxies.

In summary, using the \nuvrcorr\ vs. redshift diagram, we classify galaxies in
the COSMOS field into three subsamples, with 10,493 red, 2,347 green and 25,996
blue galaxies, respectively. We divide our sample into 8 redshift bins with bin
size $\Delta z=0.1$ within the redshift range of $z=[0.2,1.0]$. In each
redshift bin, a low mass limit $M_{\rm lim}$ was set to reduce the bias of
faint galaxies. The information of our sample is summarized in Table 1.

\subsection{Comparison sample}

Next, the properties of green galaxies are compared with those of red and blue
galaxies. To ensure that the differences shown in the comparisons are driven by
different star formation properties of different samples, but not driven by
their different mass distributions, red and blue comparison samples were made
with similar stellar mass distributions as that of the green sample. For a
green galaxy with stellar mass $M_{0}$ and redshift $z_{0}$,  matched galaxies
from red or blue samples were initially chosen. The matched galaxy meets
stellar mass $|M_{\ast}-M_{0}|<0.10$ dex at the redshift slice of
$|z-z_{0}|<0.02(1+z_{0})$ , where $M_{\ast}$ and $z$ are stellar mass and
redshift of the matched galaxy, respectively. Given that there are few blue
galaxies with stellar mass $\log(M_\ast/M_{\sun})>11.0$, the upper mass limit
of green galaxies was set to $\log(M_\ast/M_{\sun})=11.0$. The number of
matched galaxies ($N_{\rm match}$) depends on $M_{0}$ and $z_{0}$. About 98\%
green galaxies have $N_{\rm match}>15$ in both the red and blue samples. A
galaxy was then randomly selected from the matched sample for comparison of
physical properties with the green galaxy. Thus for a green galaxy sample
selected from a certain redshift bin, a blue or red comparison sample with the
same size and mass distribution was prepared. This procedure eliminated the
selection effect in the subsequent comparisons.

\section{RESULTS}

\subsection{Morphology}

We cross-matched our galaxy sample with the ZEST catalog, with an angular
diameter of $D=3.0\arcsec$. The median separation value of the first closest
counterpart is about $0.2\arcsec$. Then, a proper separation distance, $D_{\rm
separate}<0.5\arcsec$, is chosen. Considering that the ZEST catalog covers a
smaller area ($\approx 1.6$ $deg^2$) compared with the photometric catalog
($\approx 2$ $deg^2$), our final sample contains $\approx$ 25,000 galaxies with
ZEST morphological classifications.

\subsubsection{The (Gini--Asymmetry) and ($M_{\rm 20}$--Gini) diagrams}

Many previous studies found ETGs and LTGs are roughly separable on Gini
coefficient $G$ and asymmetry parameter $A$ diagram \citep{Abraham 1996, Lotz
2004, Kong 2009}. More information of the parameter definition can be found in
\citet{Conselice 2000} and \citet{Glasser 1962}. Briefly, ETGs have compact cores and regular surface
brightness distributions, thus have high $G$ and low $A$ values. For LTGs the
situation is opposite. Galaxies can be roughly classified as early-types,
late-types or irregular types based on their locations on the ($G-A$) plane
\citep{Capak 2007b}.

The $G-A$ diagrams of both green and comparison samples are plotted in Figure
4. A total of 15 blue and 15 red comparison samples were used. The blue and red
contours show the $G$ and $A$ distributions of blue and red comparison samples,
respectively. Red galaxies have high $G$ and low $A$ values, indicating that
they are compact and have regular surface brightness distributions. Indeed most
red galaxies are classified as early-type (E+S0) or early-type spirals.
Conversely, blue galaxies have low $G$ and high $A$ values and are mostly
classified as disk or irregular galaxies. The bulk of blue and red galaxies are
located in different regions on the $G-A$ plane, but some blue and red galaxies
have similar G and A values, and reside in the joint regions. Interestingly,
green galaxies are mostly distributed in the joint regions. The average $G$ and
$A$ values of green galaxies are also between those of blue and red galaxies at
each redshift bin. Note that our sample spans a redshift range of
$z=[0.2,1.0]$.  The ACS $i_{\rm 814w}$ band corresponds to 4000-6700 angstrom
in the rest-frame within this redshift range. It has been shown that the
morphological K-correction within this wavelength range is small \citep{Lotz
2004}, and the ZEST classifications are still robust \citep{Bundy 2010}.

Figure 5 shows the $M_{\rm 20}$ parameter vs. the Gini coefficient diagram.
This diagram can also effectively classify galaxies into early-types or
late-types \citep{Lotz 2006, Kong 2009, Wang 2012}. To illustrate the
robustness of morphological classification using Gini and $ M_{\rm 20}$, we
show some randomly selected galaxies on the ($G-M_{\rm 20}$) plane. The symbols
are replaced by $HST/ACS$ images (Figure 6). Generally, red quiescent galaxies
with high $G$ and low $M_{\rm 20}$ values have spheroidal morphologies, whereas
blue star-forming galaxies with low $G$ and high $ M_{\rm 20}$ values have
disk-like morphologies. Visually inspection of green galaxy images in Figure 6
suggests that most green galaxies have prominent bulges and significant disk
components. Our findings are in agreement with those of \citet{Balogh 2011} and
\citet{Mendez 2011}, that is, green galaxies are mostly at the morphologically
transiting stage between blue and red galaxies.

\subsubsection{Bulgeness}

This subsection investigates the bulge properties of disk galaxies ($T_{\rm
ZEST}$=2). Disk galaxies compromise approximately 75\% of the total galaxies.
The ZEST morphological classification assigns disk galaxies in 4 bins according
to their bulge-to-disk ratio, which is described as the bulgeness parameter
$B_{\rm ZEST}$. Since it is hard to fully separate galaxies into different
types using bivariate distributions such as Figure 4 and Figure 5, the
bulgeness parameter allows one to investigate the morphologies of green
galaxies in more detail.

Figure 7 shows the fractions of red, blue and green disk galaxies with
different $B_{\rm ZEST}$ parameter as a function of redshift. The green
galaxies were categorized into two subsamples according to their stellar mass.
Only the results of low mass galaxies at $z<0.7$ were shown in the figure
because there are few low mass green galaxies at high redshift. The error bars
were derived as the standard deviations from 15 red and blue comparison
samples. For green galaxies, the error bars were computed by 1000 bootstrap
resamplings. More information can be found in Table 2.

In the high mass bin, red galaxies are mainly composed by $B_{\rm ZEST}$=0 and
$B_{\rm ZEST}$=1 type, accounting for about 85\% to 95\%. The fractions of
these two types also show very strong redshift evolution since $z \approx 0.7$.
Our findings are in good agreement with \citet{Bundy 2010} (see Figure. 3 of
Bundy et al. 2010), suggesting that log$(M_\ast/M_{\sun})$=[10.0,11.0] red
sequence disk galaxies are more bulge-dominated and abundant since $z\approx
0.7$. Note that there are rare red pure disk galaxies within this mass range.
Blue galaxies show much stable $B_{\rm ZEST}$ and mild redshift evolution. The
fraction of blue pure disks decreases since $z \approx 0.7$. The fraction of
green galaxies classified as $B_{\rm ZEST}$=0 is significantly lower than that
of red sequence galaxies, indicating that most of them still contain
significant disk components. The jump at $z=0.25$ may due to the small sample
size (N=26) at the redshift bin. Only 5\% to 10\% green disk galaxies are
classified as pure disk systems.

Galaxies with  $\log(M_\ast/M_{\sun})<10.0$ have very different $B_{\rm ZEST}$,
in that nearly no bulge-dominated galaxies are present within this mass range
for all three samples. The $B_{\rm ZEST}$ of low mass red disks shows no
redshift evolution. However, the fraction of blue pure disk galaxies decreases
from 45\% at $z \approx 0.65$ to 20\% at $z \approx 0.25$. The $B_{\rm ZEST}$
evolution trends of green galaxies and red galaxies are similar. The most
striking feature is that, even at lower mass, the fraction of green pure disk
systems and red galaxies are similar.

In summary, the fraction of green galaxies classified as $B_{\rm ZEST}$=0 and
$B_{\rm ZEST}$=1 is significantly higher than that of star-forming galaxies,
and the fraction classified as $B_{\rm ZEST}$=3 is only about 5\% to 10\%.
Compared with red galaxies, most green galaxies still have significant disk
components, which implies that green disks are also at an intermediate stage
between blue and red disks. The lack of green pure disk galaxies suggests that
the suppression of star formation activities may be connected with galaxy bulge
formation. Considering the great $B_{\rm ZEST}$ differences between green and
blue galaxies, we disfavor the scenario that green galaxies are simply faded
blue galaxies.

\subsubsection{Compare with the Mixed Sample}

In the above two subsections, the morphological parameters of green galaxies
are found to be intermediate between those of red and blue galaxies. However,
it is still not clear whether green galaxies form a distinct population, or
their observed morphological properties can be explained by a simple mix of
blue and red galaxies. To answer this question, the morphological parameters of
green galaxies are compared with those of the ``mixed" sample. As the
morphologies of blue and red galaxies significantly differ from each other, the
morphological parameter distributions of the ``mixed" samples depend on the
ratio of blue to red galaxies ($N_{\rm blue}/N_{\rm red}$).

This part of this paper explains the process of creating a combined ``mixed"
sample, with fixed blue to red ratio of $N_{\rm blue}/N_{\rm red}=P/(1-P)$,
where $P\in(0,1)$. Firstly, for a green galaxy which with stellar mass $M_{0}$
and redshift $z_{0}$, a blue-matched sample and a red-matched sample were
formed. This was done to meet the required stellar mass $|M_{\ast}-M_{0}|<0.10$
dex at the redshift slice of $|z-z_{0}|<0.02(1+z_{0})$, where $M_{\ast}$ and
$z$ are stellar mass and redshift of the matched galaxy, respectively. Then we
generate a uniform random number $r_0$ between 0.0 and 1.0. When $r_0 \leq P$,
we randomly select a galaxy from the blue-matched sample, otherwise we select a
galaxy from the red-matched sample, to form the ``mixed" sample. A combined
blue and red galaxy sample with fixed blue to red ratio was thus formed. This
procedure also assures the ``mixed" sample and the green sample are matched in
mass and redshift.

The $G-A$ diagram of the green and ``mixed" samples is shown in Figure 8. The
green galaxies were divided into 4 redshift bins, from which mixed samples with
5 different blue-to-red ratios were created. In each panel, 15 different mixed
samples were used to minimize the random disturbance. As shown in Figure 8, low
$N_{\rm blue}/N_{\rm red}$ yields low average $A$ and high $G$ values, and the
trend is reverse at high $N_{\rm blue}/N_{\rm red}$ ratios. Green galaxies have
similar average $A$ value with the ``mixed" samples when $N_{\rm blue}/N_{\rm
red}=0.2/0.8$, whereas the average $G$ value is significantly lower. The most
reasonable compromise of average $A$ and $G$ values between green and ``mixed"
samples is observed when $N_{\rm blue}/N_{\rm red}$ is around 1.0. However, it
is clear that the $G$ scatters in green sample are \emph{significantly smaller}
than those of the ``mixed" samples, as shown in the three middle rows across
all redshift range. Green low $G$ systems have low $A$ values, which means that
they are more regular than those with similar $G$ values in the ``mixed"
samples. Therefore, a simple combination of blue and red galaxies cannot
reproduce the morphological properties of green galaxies.

Using different method, \citet{Mendez 2011} also demonstrated that green
morphological parameter distributions are  statistically different (at
2$\sigma$ level) from those of the simple mix of blue and red galaxies. Our
finding is consistent with that presented by \citet{Mendez 2011}. Thus green
galaxies most likely form a distinct population, but this phenomenon is not due
to selection effects.

\subsection{Spectral Analysis}

\subsubsection{Average Spectra of Galaxies}

The zCOSMOS provides about 10,000 galaxy spectra down to $i^{+}=22.5$,
facilitating the further exploration of spectral characteristics of green
galaxies. However, most of the spectra have very low signal-to-noise ratio
($S/N$) and are difficult to be analyzed individually. To obtain a higher $S/N$
ratio, an average spectrum for each subsample was created by stacking method
\citep{Eisenstein 2003,Shiavon 2007,Chen 2009,Shu 2012}.

Next, the photometric catalog was cross-matched with the zCOSMOS spectroscopic
sample using a $1.0\arcsec$ angular diameter, from which a sample containing
about 550 green galaxies, 1800 red galaxies and 6000 blue galaxies was
generated. The sample was divided into two redshift bins, $z=[0.2, 0.5]$ and
$z=[0.5, 1.0]$. There are few blue galaxies with stellar mass $M_\ast>
10^{11.0}M_{\sun}$, and at $z>0.5$ most red and green galaxies have stellar
mass $M_\ast> 10^{10.5}M_{\sun}$. Hence, a mass range of
log$(M_\ast/M_{\sun})=[10.0,11.0]$ for $z=[0.2,0.5]$, and
$\log(M_\ast/M_{\sun})=[10.5,11.0]$ for $z=[0.5,1.0]$ was chosen. Blue galaxies
have low $M/L$ ratio and dominate the low mass end in a flux-limited sample,
which means that a direct stacking without weight for blue sample mostly
reflects the signatures of low mass galaxies. For red sample the situation is
just reversed. To create average spectra with a same mass distribution,
galaxies were weighted as a function of mass and redshift for each subsample.
We determine the number of galaxies within mass and redshift bins (at the low
redshift bin, $ \rm \triangle log(M_\ast)$=0.34; at the high redshift bin, $\rm
\triangle log(M_\ast)$=0.25; redshift bin size is $\rm \triangle z=0.1$) both
in the full sample and in each subsample. We then weigh each spectrum by the
number of the full sample divided by the number in the subsample. This
procedure allows the breaking of mass dependence.

The average spectra of red, green and blue galaxies are shown in Figure 9. In
the low redshift bin, the average spectra range from 4600 to 6700 \AA,  and are
normalized to the average flux between 5050 \AA\, and 5100 \AA. In the high
redshift bin, the spectra are showed at shorter wavelength, including the \oii\
lines. Each average spectrum is stacked by more than 100 individual spectra
(except the green sample in the low redshift bin). Given that the mass
dependence has been minimized in the stacking procedure, these average spectra
can be treated as the ``representative spectra" of each subsample.

The red average spectra are dominated by absorption lines. The emission lines
reflecting (directly or indirectly) star formation activities, such as
H$\alpha$ and \oii\ are rare in the red spectra. The strong jump at 4000 \AA\,
and very strong Ca H+K absorption lines demonstrate that red galaxies are
dominated by old stellar populations. The metal lines, such as the Mgb and the
G-band absorption lines are most remarkable in the red spectra. Briefly, the
red average spectra are typical quiescent galaxy spectra. There are strong
emission lines in the blue spectra. The strong \oii\,, H$\alpha$ and H$\beta$
emission lines demonstrate that blue galaxies are still at active stage and
forming stars.

Interestingly, the green spectra are significantly different from the red and
blue ones. The \oii\,, H$\alpha$ and H$\beta$ emission lines are still visible,
but obviously much weaker than those of star-forming galaxies. Specifically,
the \oiii\ /H$\beta$ ratios of green spectra are higher than those of blue
spectra, confirming the result of \citet{Coil 2008}. However, in the low
redshift bin, we find the \nii\ $\lambda 6583$/H$\alpha$ ratio of the green
spectrum is still significantly lower than that of a typical AGN. The continua
of green spectra are much similar with those of red galaxies. Specifically, the
4000 \AA\, break of green galaxies is between the blue and red ones, indicating
that the average stellar age of green galaxies is between the ages of the red
and blue ones.

\subsubsection{SFH from the average spectra}

The high $S/N$ average spectra allow the extraction of galaxy star formation
history (SFH) information by comparing them with stellar population synthesis
models. This is done by comparing the average spectra in the $z=$[0.5,1.0] bin
with the STARLIGHT. \footnote{www.starlight.ufsc.br} STARLIGHT aims at fitting
an observed spectrum with a linear combination of theoretical simple stellar
populations (SSPs). The model spectrum of STARLIGHT is given by:
\begin{equation}
M_{\rm \lambda}= M_{\rm \lambda_{0}}(\sum^N_{j=1} x_{\rm j} b_{\rm
j,\lambda}r_{\rm \lambda})\otimes G(v_{\rm *},{\sigma}_{\rm *}),
\end{equation}
where $M_{\rm \lambda}$ is the model spectrum, $M_{\rm \lambda_{0}}$ is the
synthesis flux at the normalization wavelength $\lambda_0$, $x_{j}$ is the
so-called stellar population vector, $b_{j,{\lambda}}$ is the $j$th SSP
spectrum at $\lambda$, and
$r_{\lambda}\equiv10^{-0.4(A_{\lambda}-A_{\lambda_{0}})}$ represents the
reddening term. The $G(v_{*},{\sigma}_{*})$ is the line-of-sight stellar
motions which are modeled by a Gaussian distribution centered at velocity
$v_{*}$ and with a dispersion of ${\sigma}_{*}$. $N$ is the total number of SSP
models. In our work, the SSP base consisted of $N=126$ SSPs, with 6
metallicities ($Z=0.005 Z_{\odot}$, $ 0.02 Z_{\odot}$, $0.2 Z_{\odot}$,
0.4$Z_{\odot}$, $Z_{\odot}$ and $2.5Z_{\odot}$) and 21 ages (from 1 Myr to 7.5
Gyr), which were taken from evolutionary models in BC03. The maximum age
corresponded to the cosmic age at $z=$0.5. The galactic extinction law of
\citet{Cardelli 1989} with $R_{\rm V}=3.1$ was adopted. Emission lines were
masked in the fitting procedure. The STARLIGHT produced the SSP fraction, dust
attenuation $A_{\rm V}$, velocity dispersion $\sigma$, and stellar mass
$M_\ast$. Following \citet{Cid 2005}, the flux- and mass-weighed average ages
are defined, which are respectively defined as:
\begin{equation}
<{\rm log}t_{*}>_{\rm L}=\sum^N_{j=1} x_{\rm j} {\rm log} t_{\rm j};
\,\,\,
<{\rm log}t_{*}>_{\rm M}=\sum^N_{j=1} u_{\rm j} {\rm log} t_{\rm j},
\end{equation}
where $x_{\rm j}$ is the flux-weighted population vector (i.e., the
fraction of flux contributed by certain SSP), and $u_{\rm j}$ is the
mass-weighted population vector. The average metallicities $<$Z$_{\rm
L}$$>$ and $<$Z$_{\rm M}$$>$ can be derived in a similar way.

The STARLIHGT best-fit spectra as well as the SSP stellar population vectors
are presented in Figure 10. only the high redshift average spectra were fitted
because the spectra at low redshift bin did not include the 4000 \AA\ break,
which was especially important for constraining the stellar age. Spectra were
fitted up to $\lambda=4900$ \AA\, since when $z\approx 1$ spectra were
de-redshifted to $\lambda<4900$ \AA\ in the rest-frame. As demonstrated by
Figure 10, the combination of SSP spectra fits the average spectra very well.
It is worth noting that the individual output vector may be dramatically
deviated from the simulated input value. However, the average values of age and
metallicity should be more reliable, whatever they are weighted by flux or
mass. The average flux-weighted ages are 3.92, 1.81 and 1.15 Gyr for red, green
and blue spectra, respectively. A similar trend still exists when weighted in
mass. The average metallicity also shows a clear trend, wherein red galaxies
are the most metal-rich and blue galaxies are metal-poor.

The distributions of the SSP vectors are also interesting. The red spectra show
very weak emission lines and should be dominated by old stellar populations.
Interestingly, the flux-weighted vectors also show that there is no
contribution of young stellar populations ($<0.3$ Gyr) in red spectrum. The
contributions of old stellar populations ($>6.25$ Gyr) reach about 40\%. For
the green spectrum, the fraction of old populations is significantly smaller
than the red one. For the blue spectrum, there is a significant fraction ($\sim
20\%$) of very young stellar populations ($<0.01$ Gyr) and few old stellar
populations.

\subsection{Environment}

This section presents an investigation of the environment of green galaxies.
The projected $N_{\rm th}$-nearest-neighbor surface density was adopted as
local galaxy density estimator. \citet{Capak 2007b} showed that the use of
photometric redshift can also correctly characterize galaxy density in the
COSMOS field. To be consistent with previous studies \citep{Capak 2007b, Ideue
2012}, the 10th nearest neighbor was used in our analysis. The local galaxy
density was calculated as $\Sigma_{10}=11/(\pi D^{2}_{p,10})$, where $D_{p,10}$
(in Mpc) is the projected proper distance to the 10th nearest neighbor. The
projected densities was computed using $i^{+}<24.0$ galaxies. A redshift slice
centering on each galaxy with  $\pm {\Delta z}$ =0.033(1+$z$) was used,
corresponding to $\pm 3\sigma_{\rm photoz}$, where $z$ is the redshift of
central galaxy. Galaxies located near the mask regions or survey edge were
excluded from statistic. To check the reliability of the density estimator, the
sample was cross-matched with the released COSMOS field X-ray group catalog
\citep{George 2011}. The results show that the X-ray group members (which have
group member likelihood $P_{\rm mem}\geq 0.5$) mostly locate at the high
$\Sigma_{10}$ end at each redshift bin.

Figure 11 shows the $\Sigma_{10}$ histograms for blue, green and red galaxies
indicated as blue solid, green filled and red solid lines, respectively. The
$\Sigma_{10}$ distributions were averaged over 15 different comparison samples.
The error bars in the figure show the standard deviations of the 15 different
comparison samples. In addition, the $\Sigma_{10}$ histograms for the $z<1.0$
X-ray group members were plotted (shown as black shaded histograms). The global
$\Sigma_{10}$ at low redshift is higher than that at high redshift, due to
galaxy density evolution and flux limit for observation. The $\Sigma_{10}$ was
not normalized relative to the median galaxy density at each redshift bin. Blue
galaxies always prefer to reside in less dense environments, as shown in Figure
11. Conversely, red galaxies are preferentially found in dense environments.
Group members dominate the high density end at each redshift bin as expected.
However, the environmental diversities of blue and red galaxies decreases with
increasing redshift. This can be attributed to the different galaxy composition
in dense environments at different redshifts. At low redshift, galaxies in
groups or clusters are mainly red and quiescent. However, at high redshift, a
large fraction of group members still form stars and with blue colors. This is
first reported in \citet{Butcher 1978}, which is called the ``B-O" effects.

The cumulative distribution curve of each sample is shown below the histogram
panel. At $z<0.7$, green galaxies and red galaxies have similar $\Sigma_{10}$
distributions, as shown in Figure 11. Compared with blue galaxies, there is a
larger fraction of green galaxies residing in dense regions, especially at
$z<0.5$. However, in all panels at $z>$0.7, green galaxies and blue galaxies
have similar $\Sigma_{10}$ distributions. In some redshift bins, the
differences between the three cumulative lines are very subtle, which may not
be sufficient to conclude that the $\Sigma_{10}$ distributions of these
subsamples are statistically different. However, the consistent $\Sigma_{10}$
trend for green galaxies at $z<0.7$ suggests that the environment effects may
play a more important role in the formation of this population at subsequent
cosmic epoch.

To investigate the environment dependence on stellar mass, the green galaxy
sample was divided into two mass bins at $z<0.7$:
$\log(M_\ast/M_{\sun})=[10.0,11.0]$ and $\log(M_\ast/M_{\sun})<10.0$. The
$\Sigma_{10}$ distributions of the binned samples are shown in Figure 12.
Interestingly, the $\Sigma_{10}$ distributions of red and green galaxies
remarkably differ form those of blue galaxies in the low mass bin: \emph{In
comparison with blue samples, red and green samples have larger fractions of
galaxies residing in dense environments.}

However, the remarkable density distribution differences among these three
subsamples shown in the low mass bin are diluted in the high mass bin. To show
the $\Sigma_{10}$ distribution differences more statistically, we extend the
Kolmogorov-Smirnoff (KS) tests to the three distributions and summarize the
results in Table 3. The result of KS test gives small probabilities ($<1$
percent) that the $\Sigma_{10}$ of green and blue galaxies are drawn from the
same distribution at $z<0.5$ when $\log(M_\ast/M_{\sun})<10.0$. For the
$\log(M_\ast/M_{\sun})=[10.0,11]$ bin, the KS test gives larger probabilities,
in good agreement with visual inspection.

In summary, green galaxies at $z>0.7$ have similar $\Sigma_{10}$ distributions
with those of star-forming galaxies. At lower redshift, low mass green galaxies
have similar $\Sigma_{10}$ distributions with those of quiescent galaxies; they
preferentially reside in dense environments, especially at $z<0.5$. Our
findings suggest that environmental effects may be crucial for quenching star
formation activities in $\log(M_\ast/M_{\sun})<10.0$ galaxies at $z<0.7$. For
more massive galaxies, quenching does not show evident dependence on galaxy
local environment.

\section{Discussion}

\subsection{The nature of green valley galaxies}

In the above analysis green galaxies are found to have significantly different
morphological and spectral characteristics with respect to blue and red
galaxies. Their morphological properties cannot be produced by a simple mix of
blue and red galaxies, and they most likely form a distinct population.This
subsection discusses the components of the green population.

The simplest explanation for the presence of green galaxies is that they are
transiting galaxies, which are fading original star formation, on the way
towards the red sequence. The transiting scenario is required to explain the
observed redshift evolution in stellar mass functions for star-forming and
quiescent populations. Some properties of green galaxies, such as the average
Gini coefficient, the bulgeness parameter and the stacked emission lines,  are
intermediate between those of blue and red galaxies. These results can be
expected if green galaxies are at the transiting stage.

However, there are other possible ways to explain the existence of green
galaxies. One possible explanation is that green galaxies are rejuvenated red
galaxies. Red galaxies obtain gas supply either from gas-rich minor erger or
from gradual accretion, then start low level star formation activity and return
to the GV. They might move to red sequence again after the gas is exhausted.
This scenario is supported by the observation of ETGs in the FUV and NUV bands,
which reveals that a considerable fraction (about 15\% to 30\%) of ETGs have
low level recent star formation activities. Many studies favored the scenario
that the star formation in ETGs are triggered in hitherto quiescent galaxies.
Such phenomenon is termed ``rejuvenation" in the literature \citep{Rampazzo
2007, Thomas 2010, Thilker 2010, Marino 2011}. It is important to evaluate the
proportion of these sources in GV. \citet{Fang 2012} studied the extended
star-forming early-type galaxies (ESF-ETGs) in low redshift GV and estimated
that $\approx$ 13\% GV galaxies have similar UV-optical color and stellar mass
with those of ESF-ETG candidates. Another possible explanation is that galaxies
are relatively stationary in the GV because of inefficient star formation. In
this picture, galaxies will keep low level star formation for very long periods
of time due to quasi-static gas accretion from inter-galactic medium (IGM) and
never reach true passivity \citep{Salim 2012}. This so-called ``incomplete
quenching" phenomenon could be also possibly related to ''morphology quenching"
as demonstrated in \citet{Martig 2009}. In this scenario, after the cold stream
fuel star formation shut down, the further fragmentation of gaseous disks
(e.g., the ability to form stars) can be suppressed by deepening central
potential well or the declining self-gravity of the gas. This model predicts
large bulges and slow quenching in disks.

About 20\% green galaxies are classified as ETGs with ZEST. Assuming all these
ETGs are ''rejuvenated" red galaxies, they would only occupy a small fraction
of the entire green sample. In Figure 6 and Figure 7 we find the majority of
green galaxies host bulge components as well as significant disks. Their
significant disk components are more possibly inherent rather than having
formed from later gas accretion. It is difficult to distinguish those green
galaxies due to ``incomplete quenching", which prevents an accurate
proportional estimation. Nevertheless, galaxies of this category should not be
dominant in GV because they are expected to stay in GV for long periods of
time, which makes it difficult to explain the color bimodality. In summary, the
observed properties of green galaxies are consistent with the transiting
scenario.  A modest proportion of green galaxies might have different origins
and evolutional paths.

\subsection{Comparison with previous studies}

In this paper, GV was defined using dust-corrected \nuvr\ color. Because the
dust-corrected \nuvr\ color is a good tracer of sSFR, in the current paper GV
is equivalent to an intermediate sSFR (log(sSFR) $\sim -10.5$). In some
previous studies, GV is defined using optical color index such as $U-B$ or
$U-V$. Since neither $U-B$ nor $U-V$ is sensitive to intermediate sSFR, GV
galaxies defined using intermediate sSFR will degenerate with red sequence
galaxies in the $U-B$ or $U-V$ CMD. We have investigated the $U-B$ selected GV
galaxies using the criterion of \citet{Mendez 2011} and found, on the one hand,
about 75\% of them have log(sSFR)$ >-10.0$, and only about 20\% of them have
log(sSFR)$ =[-10.0,-11.0]$. On the other hand, we found about 45\% log(sSFR)$
=[-10.0,-11.0]$ galaxies are degenerated with red sequence galaxies in the
$M_{\rm B}$ versus $U-B$ diagram. One should know the differences between
optical color and sSFR GV selections when comparing our works with others.

\citet{Mendez 2011} studied morphologies of $\approx 300$ green galaxies in the
AEGIS field within the redshift range of $z=[0.4,1.2]$. They showed that green
galaxies are generally massive ($ \sim 10^{10.5} M_{\sun}$) disk galaxies with
high concentrations. The findings of \citet{Mendez 2011} are confirmed by the
current work based on a larger green galaxy sample selected from similar
redshift range. Thanks to the large green galaxy sample, the morphology of less
massive $M_{\ast}<10^{10.0} M_{\sun}$ green galaxies were also investigated.
These less massive green galaxies do not show significant increase in their
concentration since $z \approx 0.7$.

We also found that green galaxies have high \oiii$/$H$\beta$ ratio as reported
in \citet{Coil 2008}. \citet{Coil 2008} interpreted this as more AGN activity
in green galaxies as compared with blue galaxies. The samples of \citet{Coil
2008} were selected from the DEEP2 survey and mainly covered the redshift range
of $z=[0.7,1.5]$. For the galaxies at $z=[0.2,0.5]$, the \nii\ $\lambda 6583$
and H$\alpha$ emission lines are showed. As shown in Figure 9, the \nii\
$\lambda 6583$/H$\alpha$ ratio of a typical green galaxy is significantly
higher than that of a blue galaxy, yet it is still lower than a typical Seyfert
galaxy \citep{Kauffmann 2003b}. Therefore, in this work, a typical green galaxy
will be classified into the ``transition" class, using the BPT diagram
\citep{Baldwin 1981}.

\citet{Coil 2008} found that green galaxies have similar clustering amplitude
at a small scale ($<$1Mpc) as compared with blue galaxies, whereas at larger
scales green galaxies show clustering amplitude that is close to that of red
galaxies. \citet{Coil 2008} explained that green galaxies are residing in the
same dark matter halo as red galaxies, but are mainly spread in the outskirts.
Galaxy properties are expected to correlate with the parent dark matter halo,
inside which densities are evaluated on scales comparable with a typical viral
radius ($\sim$ 1 Mpc). Figure 11 shows that green galaxies at $z>0.7$ have
similar $\Sigma_{10}$ distributions with blue galaxies. In comparison with red
galaxies, green galaxies tend to prefer lower density environments, which is
also in agreement with the findings of \citet{Coil 2008}. \citet{Balogh 2011}
reported the existence of a large green population in X-ray group. We will
compare our results with those of \citet{Balogh 2011} and discuss in detail in
Section 5.5.

\subsection{Quenching connected with bulge formation}

About 90\% of green disk galaxies have bulge components. Specially, the
fraction with prominent bulge is about 60\% to 70\% in the green disk sample,
and only $\approx$ 35\% in the blue disk sample. \emph{The presence of bulge in
the majority of green galaxies suggests that star formation quenching in this
population possibly accompanies or has connection with bulge formation}.

Recent studies have found that quiescent galaxies are mostly associated with
the presence of bulges. \citet{Bundy 2010} found that passive disks typically
have Sa--Sb morphological types and large bulges, based on the COSMOS data;
\citet{Kauffmann 2012} found there are clear thresholds in bulge-to-disc ratio
and in stellar surface density that demarcate the location of quiescent
galaxies. They proposed that the processes associated with bulge formation play
an important role in depleting the gas in galaxies; \citet{Bell 2012} studied a
sample from CANDELS and found that most quiescent galaxies host prominent
bulges, and many of them have significant disks. They argued a prominent bulge
(and perhaps associating with a supermassive black hole) is an important
condition for quenching star formation on galactic scale over the last 10 Gyr.
Our findings confirm that both green and red galaxies tend to have prominent
bulges.

\citet{Cheung 2012} searched for traces of possible quenching processes through
galaxy structural parameters and found the central surface stellar mass density best discriminates between red sequence and blue cloud. They found that red sequence bulges are roughly twice as massive as blue cloud bulges at the same stellar mass. Their results suggest that the innermost structure of galaxies is most physically linked with quenching. Some possible quenching mechanisms are
discussed in \citet{Cheung 2012}. However, the ways by which these mechanisms
relate to bulge-buildup is still unclear.

The current study finds that a small fraction (about 5\% to 10\%) of green disk
galaxies does not harbor identified bulge components. This finding is
consistent with that of \citet{Mendez 2011}, who reported that 12\% of their
green galaxies have $B/T$=0. We stress that there is no clear boundary
separating green and blue galaxies in the morphological parameter planes. These
bulgeless green disks must have their star formation suppressed by other
mechanisms without forming galactic bulges.

\subsection{Quenching connected with AGNs}

There are some recent indications that active galactic nucleus (AGN) inhabits
GV, both from X-ray surveys \citep{Nandra 2007, Coil 2009} and optical
line-ratio diagnostics \citep{Salim 2007}. However, it is still unclear whether
there are direct relations between AGNs and star formation quenching process in
green galaxies. One the one hand, some studies found observational evidence
that a powerful AGN can drive high speed outflow expelling the interstellar
mediums, and will suppress the star formation of its host galaxy in a very
short period of cosmic time \citep{Feruglio 2010, Sturm 2011,Page 2012}. On the
other hand, studies based on large samples reported that AGNs have complicate
relations with star formation activities, but may not be directly responsible
for quenching star formation \citep{Bundy 2008, Alonso 2008, Aird 2012}. Until
now, despite recent progress, how AGNs regulate the star formation and shape
galaxy global properties is a highly debatable.

To understand the role of AGN feedback, it is important to compare the
properties of AGN host galaxies with those of inactive galaxies. While
considering the fact that when a galaxy hosts an AGN, its observed SED is a
superposition of an AGN and the host component. Thus one must subtract the AGN
contribution from the total observed flux before deriving the correct color and
mass of the host galaxy. Studying the link between AGNs and their hosts is a
key question in astrophysics.  However, this is beyond the scope of the current
work. In this paper, we will just briefly discuss the recent progresses in this
field.

\citet{Xue 2010} found moderate-luminosity AGN hosts apparent no color
bimodality on the CMD. When matched in mass, AGN hosts lie in the same region
on the CMD as inactive galaxies. They suggested that the presence of
moderate-luminosity AGN activity does not have a significant effect on the
galaxy color. \citet{Bongiorno 2012} explored the host galaxy properties of
$\approx$ 1700 AGNs in the COSMOS field, and found AGN hosts are mainly
massive, red galaxies. They found no conclusive evidence suggesting that AGNs
have powerful influence on the star-forming properties of their host galaxies.
\citet{Rosario 2013a} studied the properties of X-ray selected AGN hosts in the
Cosmic Assembly Near-IR Deep Legacy Survey and found, the colors, color
gradients and stellar population properties of AGN hosts are similar with those
of inactive galaxies with the same stellar mass. However, in a most recent
study, using far-infrared observations of the two Chandra Deep Fields,
\citet{Rosario 2013b} found the UV-to-optical colors of AGNs are consistent
with equally massive inactive galaxies at same redshift, whereas their FIR
distributions are statistically similar with those of star-forming galaxies.

Understanding the importance of AGN feedback requires a complete AGN sample
spanning a wide luminosity range. The AGN samples used in previous studies are
selected by X-ray, spectral indices or IR SEDs. As suggested in many previous
studies, a significant fraction of obscured AGNs will not be identified in
insufficient deep X-ray surveys. The IR AGN selection criteria are relatively
insensitive to dust obscuration. However, as suggested in \citet {Mendez 2013},
the IR selection appears to be biased in selecting high luminosity AGNs. We
have used the mid-IR AGN selection criteria from \citet{Donley 2012} to select
AGNs from our galaxy sample. 90 AGNs were selected with the IRAC photometry,
and most of them are blue. However, this must be a small fraction of the actual
number, which is likely owing to the relatively more shallower photometry in
IRAC bands with respect to the optical bands. To summary, whether AGN feedback
is responsible for quenching cannot be well constrained in our study.

\subsection{Quenching connected with environmental effects}

Galaxies in dense regions are influenced by environmental conditions, such as
galaxy interactions, tidal forces due to the cumulative effect of many weaker
encounters, and gas striping due to the interaction with IGM, which can affect
(more or less) their evolutional paths. A great deal of works support that the
dense environments are especially important for gas stripping/removal in LTGs
and turning them into ETGs \citep{Boselli 2006, Cooper 2006, vandenBosch
2008,Smith 2010,vanderwel 2010, Weinmann 2010}.

Our findings show that green galaxies exist in various environments, indicating
that environmental effect is not the sole explanation responsible for quenching
star formation in this population. In Section 4.1.2, most green disks are found
to host significant bulges as well as disk components. These green disks are
likely the progenitors of S0 or passive spirals. \citet{Bundy 2010} found that
passive spirals in the COSMOS filed harbor large bulges, but not confined to
dense environments. However, less massive green galaxies
($M_{\ast}<10^{10.0}M_{\sun}$) at $z<0.7$ prefer dense environments, strongly
suggests that the environmental conditions may play an important role on
quenching at the later cosmic time.

Although both the morphological evolution \citep{Kovac 2010} and star formation
activity are influenced by environment, the dominated mechanism affecting
galaxy evolution in dense environments is still debatable. Because there are
only 1-2 clusters in the COSMOS field, strong ram pressure stripping, which
works in cluster cores, will not play a dominant role in quenching in this
study. In a group environment, frequent galaxy interactions are expected to be
especially efficient in influencing galaxy evolution. Tidal interactions can
trigger star formation \citep{Lambas 2003, Nikolic 2004, Li 2008}, which could
be a factor in the transformation of blue, star-forming galaxies to red,
quiescent galaxies.\citep{Kenneth 2011, Patton 2011}.

Comparing the merger rate between blue, green and red galaxies may be helpful
in assessing the role of merger on the formation of green galaxies. However,
accurate measurement of merger fraction requires spectroscopic redshifts and
very high sampling rate. The zCOSMOS survey has low sampling rate ($<0.1$) for
galaxies with close companions ($D_{\rm separation} <5.0 \arcsec$), which
restricts an accurate merger rate measurement. We note that the ways by which
mergers shape galaxy properties have been widely discussed in literature.
\citet{Perez 2009} suggested that, at intermediate densities, close pairs could
have experience more rapid transition onto the red sequence than isolated
galaxies. \citet{Ellison 2010} found that there is a clear enhancement of
bulge/total ratio (B/T) when pairs have small separations ($<30$ $h^{-1}$ kpc).
They interpreted this as the signature of central star formation in interacting
pairs. \citet{Mendez 2011} found most green galaxies in the AEGIS field are not
classified as mergers, and their merger rate is even lower than that of blue
galaxies. \citet{Alonso 2012} studied galaxy pairs in SDSS groups and found,
the CMD of groups/clusters pairs consists of a clear excess of extremely blue
and red galaxies with respect to that of the control sample. Most of these
studies suggest that mergers enhance star formation activities, and the
interacting galaxies evolve more rapidly than their isolated counterparts. If
this is the case, then the interacting galaxies will pass their GV phase in a
shorter time than isolated galaxies. Thus, the results shown in \citet{Mendez
2011} and \citet{Alonso 2012} are natural.

In galaxy groups, very weak ram pressure can strip gas, thereby gradually
shuting off star formation. This process is known as ``strangulation".
``Harassment", which is known as the cumulative effects of many weak encounters
and tidal interactions, also works in a group environment. These two mechanisms
are thought to be especially important to the evolution of low mass galaxies
\citep{Barazza 2002, Haines 2006, Haines 2008}. As Compared with merger or
strong galaxy-galaxy interactions, strangulation and harassment mainly act on
gas content and will not modify galaxy structure directly. Violent merging of
two spiral galaxies is most likely to produce a bulge-dominated ETG.
Simulations have showed that the remnants of strangulation should have similar
structures with those of disk galaxies \citep{McCarthy 2008}, which may help in
interpreting the existence of bulgeless green disks. The lack of low mass
bulge-dominated systems in both red and green disks (see Figure 7) suggests
that, in a group environment, strangulation and harassment are the most likely
mechanisms responsible for quenching.

The current study find that green galaxies reside in similar density
environments as blue galaxies at $z>0.7$. \citet{Balogh 2011} targeted 7 galaxy
groups in the COSMOS field and claimed that they discovered a large fraction of
transiting galaxy population in X-ray groups at $0.85<z<1.0$. However, their
sample size is much smaller (about 100 galaxies in total, and about 20 of them
are green galaxies) than ours. They used the k-corrected ($V-z$) color index to
define GV, whereas we use the dust-corrected \nuvr\ color. Thus the composition
of green galaxies are quite different in these two works, as also discussed in
Section 5.2. In the left panel of Figure 13 we show the \nuvrcorr\ and
$(V-z)^{0.9}$ for galaxies at $0.8<z<1.0$, with the green galaxy criterion of
\citet{Balogh 2011} shown in vertical dashed lines. It is obvious that these
two criteria select different galaxies. The GV defined by $(V-z)^{0.9}$ color
is contaminated by many dusty blue galaxies. Galaxies are more efficiently
separated into two main categories (blue star-forming and red quiescent) using
dust-corrected \nuvr\ color. In addition, our GV definition is much stricter
than that of \citet{Balogh 2011}, which can exclude the contaminants of
star-forming or quiescent galaxies more efficiently. We suggest the
dust-corrected \nuvr\ color should be more reliable for selecting truly
transiting galaxies.

Following \citet{Balogh 2011}, the $(V-z)^{0.9}$ color distribution of X-ray
group galaxies was compared with that of field galaxies, to check whether there
was a large amount of transiting galaxies at $0.8<z<1.0$. The group members
were selected as galaxies with $P_{\rm mem}\geq 0.1$ in the X-ray group catalog
of \citet{George 2011}. A low $P_{\rm mem}$ limit was applied in order to keep
a high group member completeness. As illustrated in \citet{George 2011}, a
$P_{\rm mem}\geq 0.1$ member selection limit maintain 95\% true group members,
but brings in about 35\% field contaminants. The field galaxies were selected
as those residing in sparse environments (which have log$(\Sigma_{10})<0.5$ and
$P_{\rm mem}=0.0$ when cross-matched with the X-ray group catalog). The results
are shown in the middle and right panels in Figure 13. As shown in Figure 13,
the $(V-z)^{0.9}$ color distribution of field galaxies is very similar with
that reported in \citet{Balogh 2011}. Given that the purity of the selected
group members is about 65\%, there is a need to determine if "fake" group
members will distort the color distribution of "true" group members.

To do this the color distribution of ``interlopers" was subtracted from that of
the selected group members. The purity and completeness are formally given by:
\begin{equation}
p=\frac{N_{\rm selected}-N_{\rm interlopers}}{N_{\rm selected}}, \hspace{1.0cm}
c=\frac{N_{\rm ture}-N_{\rm missed}}{N_{\rm true}},
\end{equation}
where $N_{\rm true}=N_{\rm selected}+N_{\rm missed}-N_{\rm interlopers}$. There
are 443 galaxies with $P_{\rm mem}\geq 0.1$, thus the number of ``fake" group
members is about $443\times(1-0.65)\simeq 155$. These 155 interlopers were
assumed to have color distributions as those of field galaxies. The
$(V-z)^{0.9}$ color distribution of the interlopers-corrected group members is
shown as a solid histogram in the middle panel of Figure 13. For comparison,
the dust-corrected \nuvr\ distribution was plotted in the right panel of Figure
13. We find that group members tend to have redder color with respect to field
galaxies, whatever termed by $(V-z)^{0.9}$ or \nuvr. The $(V-z)^{0.9}$ color
distribution of group member is perfectly consistent with that reported in
\citet{Balogh 2011}. Compared with the $(V-z)^{0.9}$ color distribution of
field galaxies, there is a remarkable excess of green and red galaxies located
in groups at the expense of blue galaxies. The excess of galaxies residing in
GV is not seen after dust correction. Most of the $M_{\ast}>10^{10.1}M_{\sun}$
green galaxies defined by $(V-z)^{0.9}$ color fall in the massive star-forming
category when using the dust-corrected \nuvr\ classification. In summary, we
successfully reproduce the results of \citet{Balogh 2011}. However, after dust
correction, there is no excess of $M_{\ast}>10^{10.1}M_{\sun}$ group galaxies
residing in GV with respect to the field.

\subsection{The formation of red galaxies}

Previous works have suggested a significant growth of red sequence galaxies
since $z \approx 1.0$. Recently, \citet{Moustakas 2013} measured the evolution
of stellar mass function form $z=0-1$ using the PRIMUS and SDSS data. On the
one hand, they found the space density of massive star-forming galaxies
declines by 54\% since $z \approx 1$. On the other hand, the most massive
quiescent galaxies are largely in place since $z\approx 1$. The build-up of red
sequence occurs significantly at intermediate and low mass, where they found a
factor of 2-3 increase in the number density of $10^{9.5}-10^{10}M_{\sun}$
galaxies since $z \approx 0.5$. In this study, our findings suggest a
significant fraction of these low mass red sequence galaxies are likely
quenched due to environmental effects.

We find that the build-up of red sequence is strongly mass-dependent. At
$z>0.7$, the majority of red galaxies are massive, with $\log
(M_{\ast}/M_{\sun})>10.0$. Less massive galaxies join the red sequence at
$z<0.7$. The age--downsizing, according to which less massive galaxies contain
younger stellar population, have been prevalently found in studies on ETGs
\citep{Thomas 2005, Zhu 2010, Pan 2012}. Some models have been proposed to
explain the observation \citep{Faber 2007, Pozzetti 2010}. \citet{Pozzetti
2010} proposed a scenario wherein intermediate-mass galaxies decrease their
star formation activities to intermediate values gradually, and finally to the
values similar to of those quiescent galaxies, by means of the exhaustion of
either the gas reservoir or cold gas accretion, or by quenching due to AGN
feedback. For the schematic of this scenario, one can refer to Figure. 21 of
\citet{Pozzetti 2010}. In this scenario, massive star-forming galaxies have
shorter quenching time scale. Finally, they evolve to red sequence on a time
scale of $1 \sim 2$ Gyr, and undergo a dynamical transformation into spheroidal
galaxies.

Our findings suggest that quenching relates to both mass and local density
environment. However, which process (mass quenching and environment quenching)
plays the dominant role is still unclear. \citet{Peng 2010} have taken an
empirical approach to identify how mass and environment affect the ``quenching"
process at different cosmic epochs. They found that the effects of mass and
environment are completely separable to $z \sim 1.0$. The physical origins of
environment quenching have been discussed in Section 5.5, but the actual
physical mechanism leading to mass quenching remains largely unknown.
\citet{Peng 2010} summarized the dominant mechanism responsible for quenching
as follows: at high masses $(M_{\ast}>10^{10.5}M_{\odot})$, mass quenching
always plays a dominant role. At lower masses, merger quenching works at
$z>0.5$ and environment quenching works at $z<0.5$. Our results can be well
explained by \citet{Peng 2010} model, although the mass and redshift thresholds
separating the dominant mechanism are slightly different in these two works.

\section{Summary}

In this paper, a sample of $\approx 2350 $ green galaxies was constructed using
dust-corrected \nuvr\ color in the COSMOS field, after which their properties
were compared with those of mass-matched blue and red samples at $0.2<z<1.0$.
The findings are summarized as follows:

1. Green galaxies mostly locate in the joint region of blue and red galaxies in
the galaxy structural parameter diagrams, both on the (Gini--Asymmetry) plane
and the $(M_{\rm 20}$--Gini) plane.

2. Red disk galaxies mostly host large or intermediate bulge components,
accounting for about 80\% to 90\%. The fraction of green galaxies with large or
intermediate bulges accounts for about 60\% to 70\%, which is much higher than
that of blue galaxies (about 35\%). Only 5\% to 10\% green spirals are
classified as pure disk systems. Most green galaxies still harbor significant
disk components.

3. The stacked spectra of green galaxies have intermediate emission lines, such
as \oii\,, H$\beta$ and H$\alpha$. The SSP decomposition of the stacked spectra
shows that green galaxies have small fraction of young stellar populations,
whereas the old populations account for very large fractions. The average
metallicity of green galaxies is higher than that of blue galaxies and lower
than that of red galaxies.

4. The environments of green galaxies were explored using the local galaxy
density $\Sigma_{10}$. At $z>0.7$, green galaxies tend reside in similar
density environment with that of blue galaxies; whereas at lower redshift, a
large proportion of $M_{\ast}<10^{10.0}M_{\odot}$ green galaxies reside in
dense environment, which is significantly different with that of
$M_{\ast}<10^{10.0}M_{\sun}$ blue galaxies.

Our findings support that the physical properties of green populations are
consistent with those of the transiting populations between blue star-forming
galaxies and red quiescent galaxies. We are able to reproduce the color
distribution for X-ray groups members as reported in \citet{Balogh 2011} at
$0.8<z<1.0$. However, no excess of green galaxies is found in groups with
respect to the field after dust-correction. Our findings support the galaxy
evolution model of \citet{Peng 2010}.

\acknowledgments

We thank the referee for thoughtful comments and insightful suggestions
that improved our paper greatly.  This work is supported by the National
Natural Science Foundation of China (NSFC; Nos. 11225315 and 11203023),
and the Chinese Universities Scientific Fund (CUSF) and the Specialized
Research Fund for the Doctoral Program of Higher Education (SRFDP;
No. 20123402110037).  We thank the COSMOS team to make their excellent
data publicly released, Dr. Guanwen Fang and Dr. Tao Wang for valuable
discussion.

\newpage

\begin{table}
\begin{center}
\caption{Low mass limit and galaxy number of blue, green and red sample
in each redshift bin}
\begin{tabular}{@{}crrrr}
\tableline\tableline
redshift &mass limit &$\rm N_{blue}$&$\rm N_{green}$ &$\rm N_{red}$\\
    &log$(M/M_{\sun}$)& &\\
\tableline
0.2-0.3   &8.4  &3101    &156  &842   \\
0.3-0.4   &8.7  &5522    &405  &1824   \\
0.4-0.5   &8.9  &3763    &303  &1072  \\
0.5-0.6   &9.1  &3327    &286  &794   \\
0.6-0.7   &9.4  &3751    &453  &1599  \\
0.7-0.8   &9.5  &3198    &189  &1157  \\
0.8-0.9   &10.0 &2068    &345  &1927  \\
0.9-1.0   &10.3 &1265    &210  &1278  \\
\tableline
\end{tabular}
\end{center}
\end{table}

\begin{table*}
\begin{center}
\caption{$B_{\rm ZEST}$ for disk galaxies.}
\begin{tabular}{@{}crrrrr}
\tableline\tableline
 redshift &mass range&galaxy & $B_{\rm blue}^a $&$B_{\rm green}$&$B_{\rm
 red}$\\
 &log$(M/M_{\sun})$&number& & &\\
\tableline
0.2-0.3 &10.0-11  &26   &[ 4,11, 9, 2]  &[ 3,11, 9, 3]   &[16,19, 1, 0]\\
0.3-0.4 &10.0-11  &94   &[12,38,32,12]  &[27,41,25, 1]   &[50,40, 4, 0]\\
0.4-0.5 &10.0-11  &99   &[12,37,37,13]  &[23,52,22, 2]   &[46,50, 3, 0]\\
0.5-0.6 &10.0-11  &108  &[10,38,37,23]  &[26,56,20, 6]   &[43,55, 9, 1]\\
0.6-0.7 &10.0-11  &191  &[17,63,64,47]  &[41,88,47,15]   &[57,114,17,3]\\
0.7-0.8 &10.0-11  &85   &[ 9,25,27,23]  &[11,46,21, 7]   &[26,49, 9, 1]\\
0.8-0.9 &10.0-11  &170  &[22,52,55,41]  &[29,80,39,22]   &[49,98,22, 1]\\
0.9-1.0 &10.0-11  &90   &[10,29,30,21]  &[14,43,26, 7]   &[22,61, 6, 1]\\
\tableline
0.2-0.3 &$<$10.0  &65   &[ 2,15,34,14]  &[ 2,27,33, 3]   &[ 5,30,36, 4]\\
0.3-0.4 &$<$10.0  &123  &[ 3,28,62,30]  &[ 9,46,58,10]   &[ 8,72,41, 2]\\
0.4-0.5 &$<$10.0  &59   &[ 2,15,25,17]  &[ 1,28,25, 5]   &[ 6,31,21, 1]\\
0.5-0.6 &$<$10.0  &31   &[ 1, 6,13,11]  &[ 1,13,14, 3]   &[ 5,16,10, 2]\\
0.6-0.7 &$<$10.0  &50   &[ 1,10,17,22]  &[ 1,20,23, 6]   &[ 2,28,17, 3]\\
\tableline
\end{tabular}
\tablenotetext{a}{Number of disk galaxies classified as $ B_{\rm
ZEST}=[0,1,2,3]$. For red and blue galaxies , the number is averaged over 15
different comparison samples, respectively.}
\end{center}
\end{table*}

\begin{table}
\begin{center}
\caption{Results of KS statistical tests applied to the $\Sigma_{10}$
  distribution in green galaxies and comparison sample.}
\begin{tabular}{@{}crrrr}
\tableline\tableline
 redshift &mass range &galaxy number&Probability &Probability\\
 &log$(M/M_{\sun})$& &$KS_{\rm green-red}$ &$KS_{\rm green-blue}$\\
\tableline
0.2-0.3   &$<$10.0  &99    &0.1634  &5.35E-5   \\
0.3-0.4   &$<$10.0  &201   &0.1913  &2.6E-14   \\
0.4-0.5   &$<$10.0  &106   &0.4676  &0.00013   \\
0.5-0.6   &$<$10.0  &66    &0.0389  &0.01625   \\
0.6-0.7   &$<$10.0  &74    &0.1497  &0.08104   \\
\hline
0.2-0.3   &10.0-11  &33    &0.2636  &0.86874   \\
0.3-0.4   &10.0-11  &138   &0.6733  &0.00331   \\
0.4-0.5   &10.0-11  &137   &0.0959  &0.65454   \\
0.5-0.6   &10.0-11  &150   &0.0514  &0.12042   \\
0.6-0.7   &10.0-11  &261   &0.6705  &8.27E-5   \\
\tableline
\end{tabular}
\end{center}
\end{table}

\newpage

\begin{figure}
\centering
\includegraphics[width=0.9\columnwidth,angle=0]{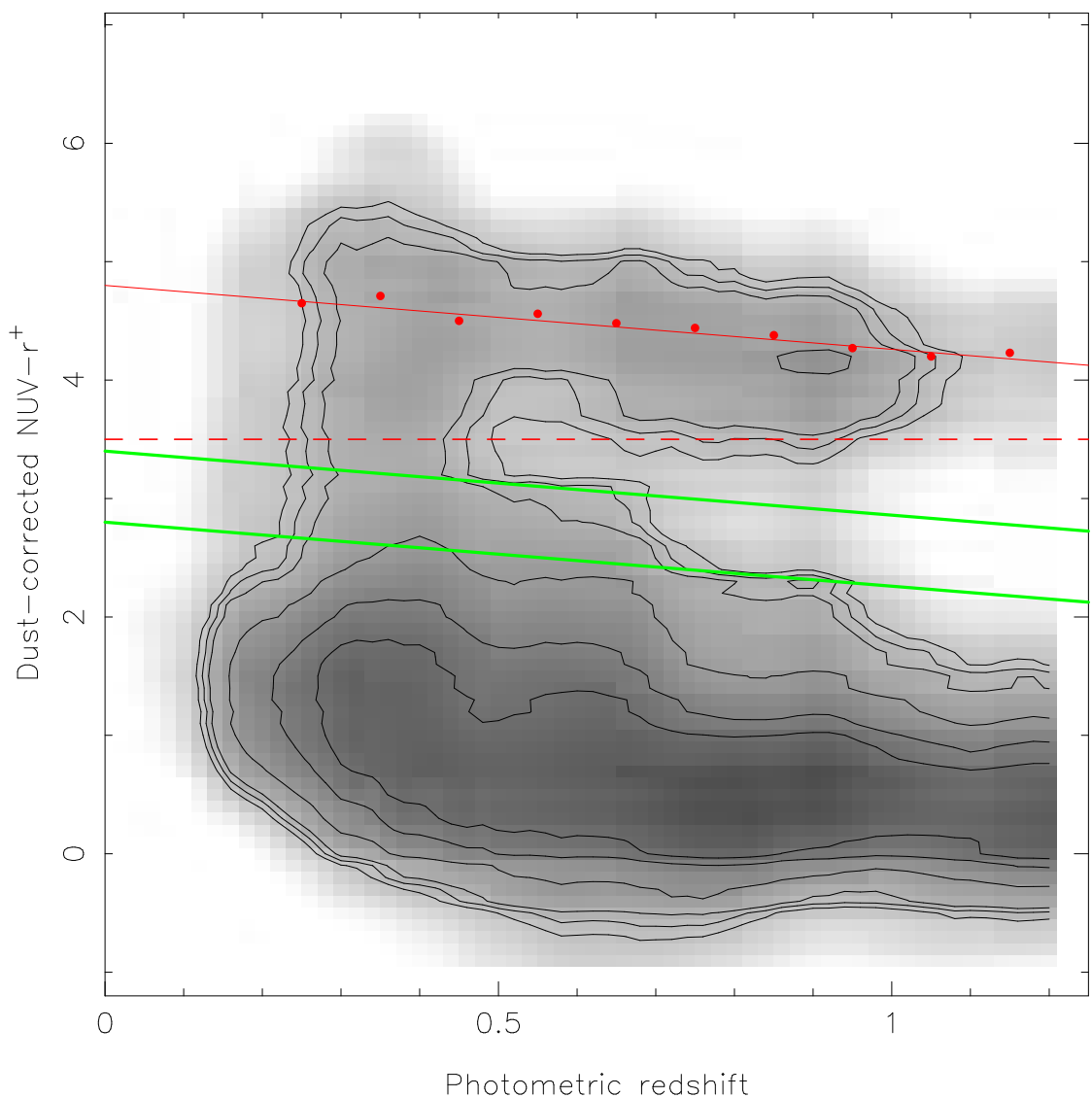}
\caption{Rest-frame \nuvrcorr\ color as a function of redshift for galaxies
with $i^{+}<24.0$ and $M_{\ast}>10^{8.0}M_{\sun}$ at $z<1.2$. The grayscale
represents the galaxy number density. The contours are plotted to clarify the
bimodal color distribution. The red solid line is the linear fit of the medium
color of ``quiescent galaxies". Red dashed line presents the color cut utilized
to separate ``quiescent" galaxies, as quoted in Ilbert et al. (2010). The red
solid circles mark the medium color for ``quiescent" galaxies within the
redshift range of $0.2<z<1.2$. The two green solid lines show the selection
criteria of green galaxies. }
\end{figure}

\begin{figure*}
\centering
\includegraphics[width=0.9\textwidth, angle=0]{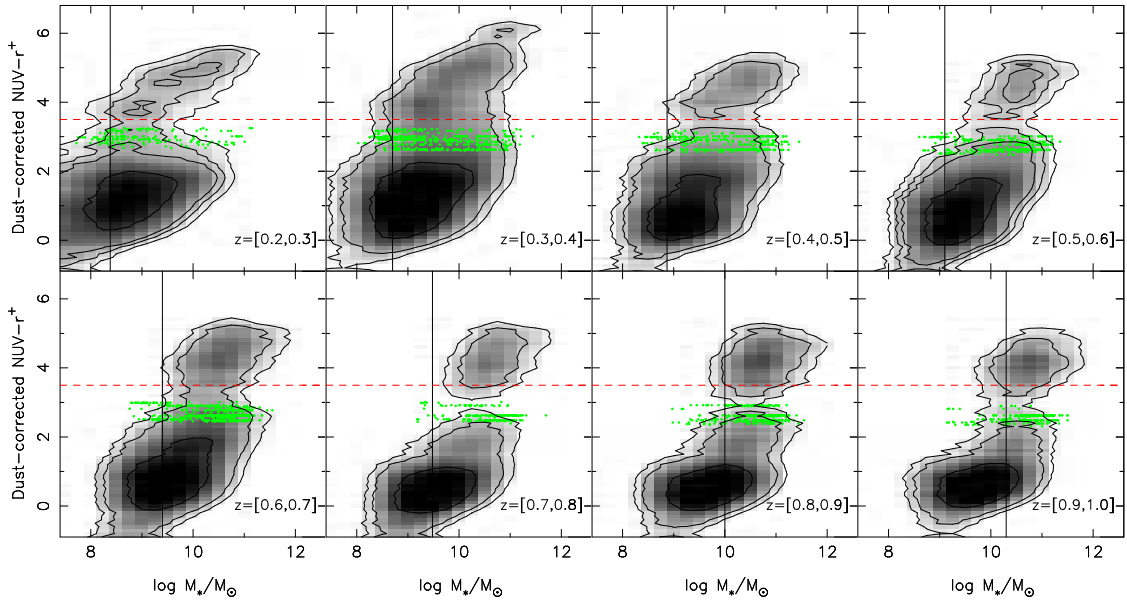}
\caption{
\nuvrcorr\ color as a function of stellar mass for galaxies with
$i^{+}<24.0$ within the redshift range of $z=[0.2,1.0]$. The grayscale
represents galaxy number density. Red dashed line denotes the color cut of
\nuvrcorr\ $=$ 3.5. Green dots are green galaxies that meet our selection
criterion. Solid vertical line shows our low mass limit for each redshift bin.}
\end{figure*}

\begin{figure*}
\centering
\includegraphics[width=0.9\textwidth, angle=0]{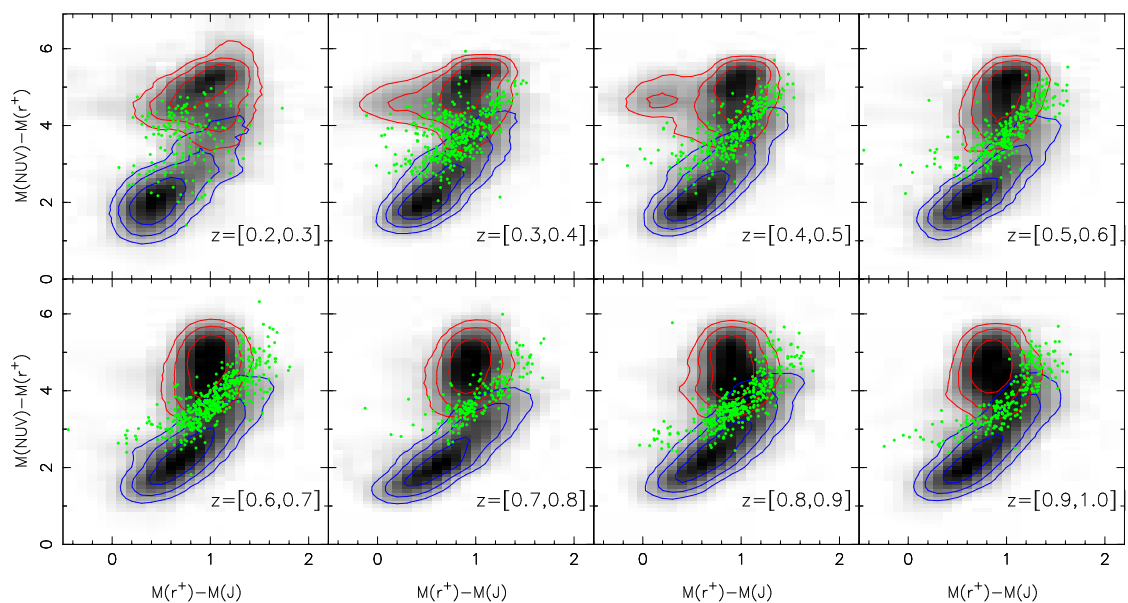}
\caption{
 Rest-frame $M{\rm (NUV)}-M{\rm (r^{+})}$ vs. $M{\rm (r^{+})}-M{\rm
(J)}$ (no dust-corrected) from $z=0.2$ to $z=1.0$, for galaxies above low mass
limit. The grayscale represents  galaxy number density. Red and blue contours
include 95\%, 80\% and 60\% of red and blue galaxies, respectively. Green dots
denote green galaxies in our sample.}
\end{figure*}

\begin{figure*}
\centering
\includegraphics[width=0.9\textwidth,angle=0]{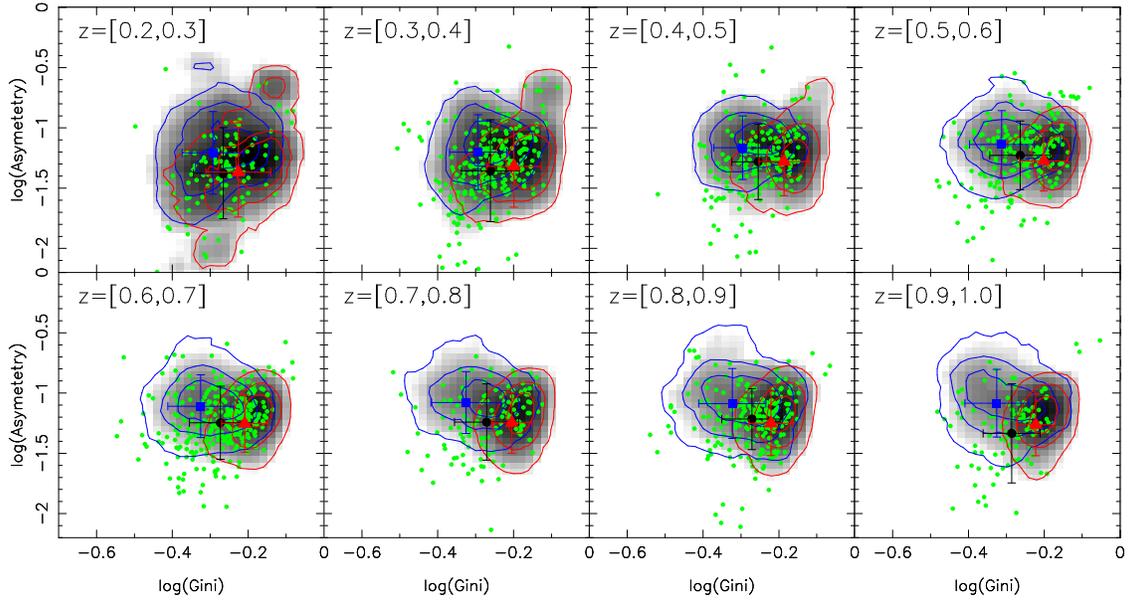}
\caption{ The Gini coefficient ($G$) vs. asymmetry ($A$) diagram. Green
galaxies are denoted as green dots. The numbers of the red and blue compared
sample are 15 times as that of the green sample. The grayscale represents
galaxy number density of comparison sample. The red and blue contours enclose
30\%, 60\% and 90\% of the red and blue galaxies, respectively. Blue square,
black circle and red triangle mark the average Gini coefficient and asymmetry
of the blue, green and red sample, respectively. The error bars are the
standard deviations of $G$ and $A$ distributions for each sample.}
\end{figure*}

\begin{figure*}
\centering
\includegraphics[width=0.9\textwidth, angle=0]{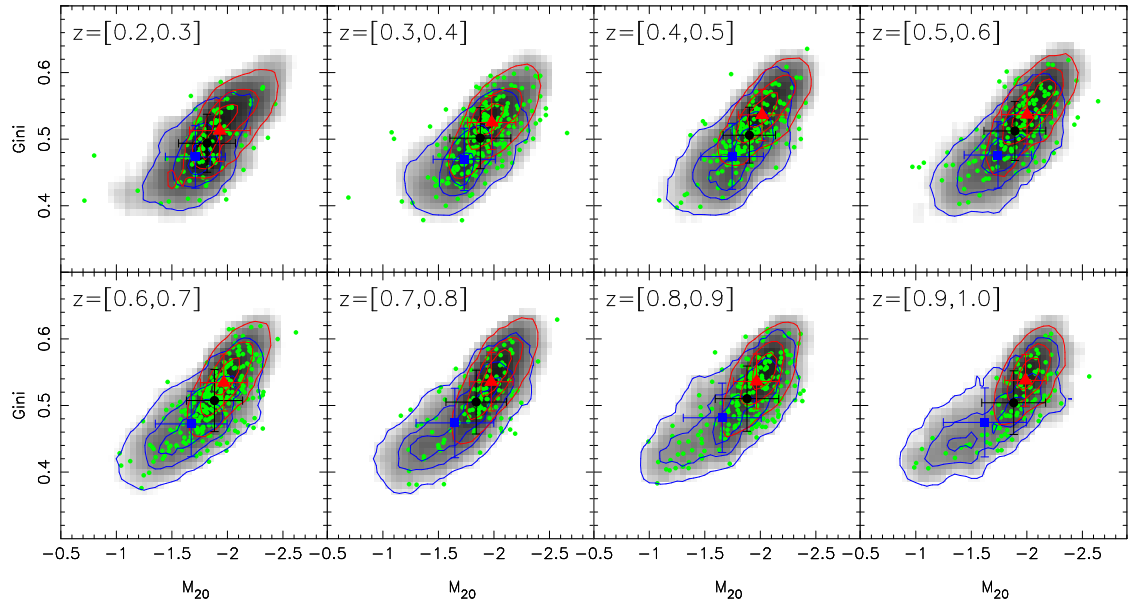}
\caption{The $M_{\rm 20}$ vs. Gini diagram. Symbols are the same as
in Figure 4.}
\end{figure*}

\begin{figure}
\centering
\includegraphics[width=0.9\columnwidth, angle=0]{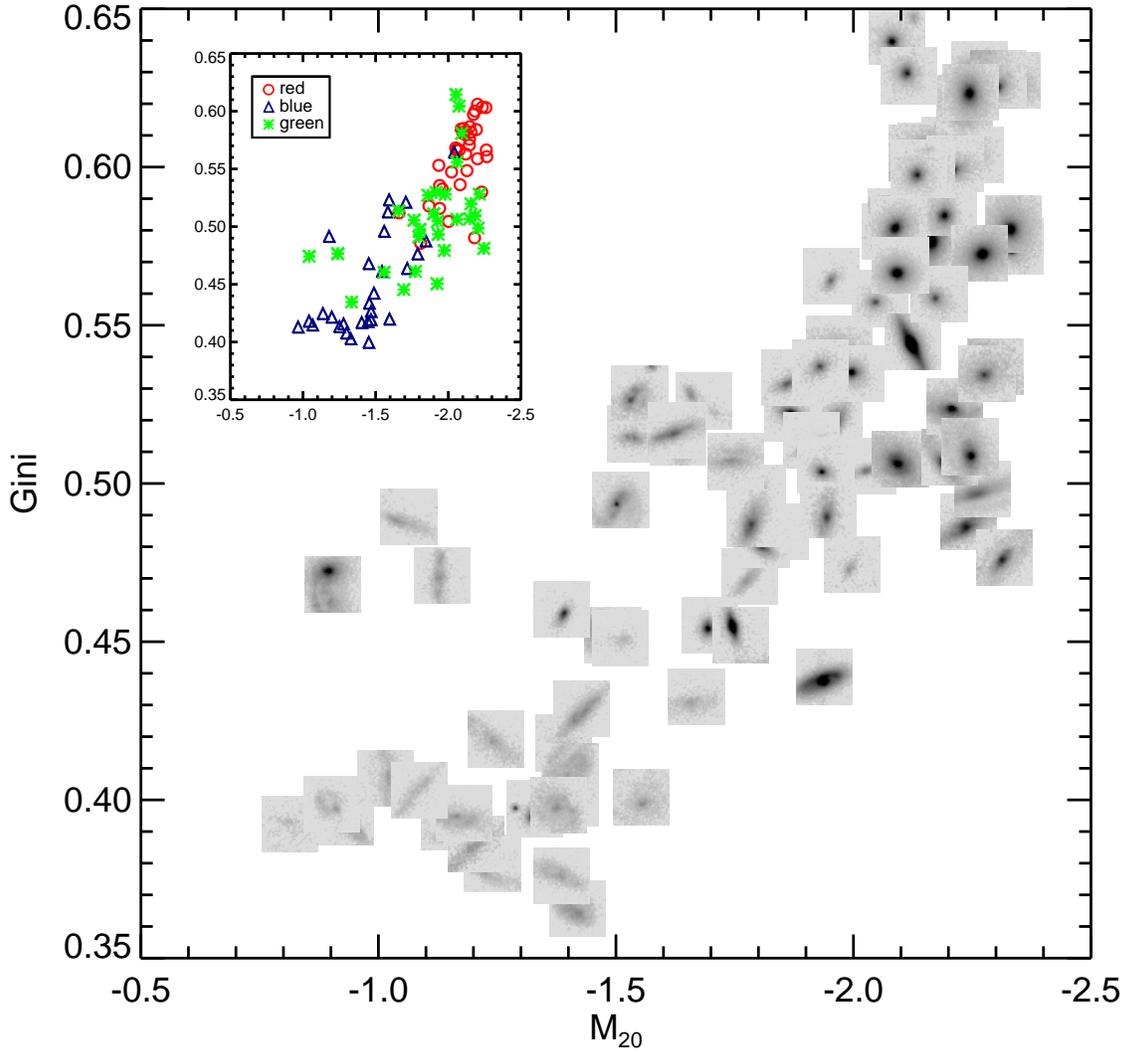}
\caption{
The $M_{\rm 20}$ vs. $G$ diagram for some randomly selected galaxies
at $0.6<z<0.7$. Red, green, blue galaxies are denoted with red circles,
green crosses and blue triangles, respectively. The HST/ACS postage-stamp
images for each subsample represent morphologies for galaxies in various
locations of this diagram.}
\end{figure}

\begin{figure*}
\centering
\includegraphics[width=0.9\textwidth, angle=0]{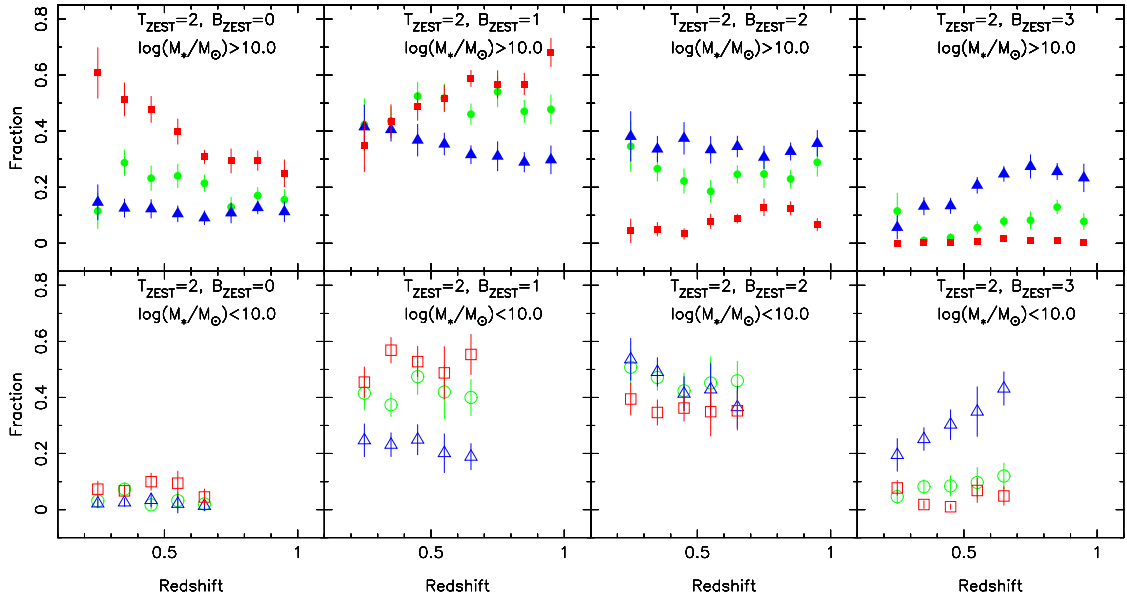}
\caption{ The fraction of red, blue and green disk galaxies with different
$B_{\rm ZEST}$ parameter as a function of redshift. For a certain subsample,
the fraction is defined as the disk galaxy number with a certain $B_{\rm ZEST}$
in the subsample divided by the total disk galaxy number in the subsample in
that redshift bin. Blue galaxies are indicated in blue triangles, green
galaxies in circles, and red galaxies in squares. For blue and red compare
samples, the error bars are the 1 $\sigma$ standard deviations of 15
resamplings. The error bars of green galaxies are derived from 1000 bootstrap
resamplings. }
\end{figure*}

\begin{figure*}
\centering
\includegraphics[width=0.9\textwidth,angle=0]{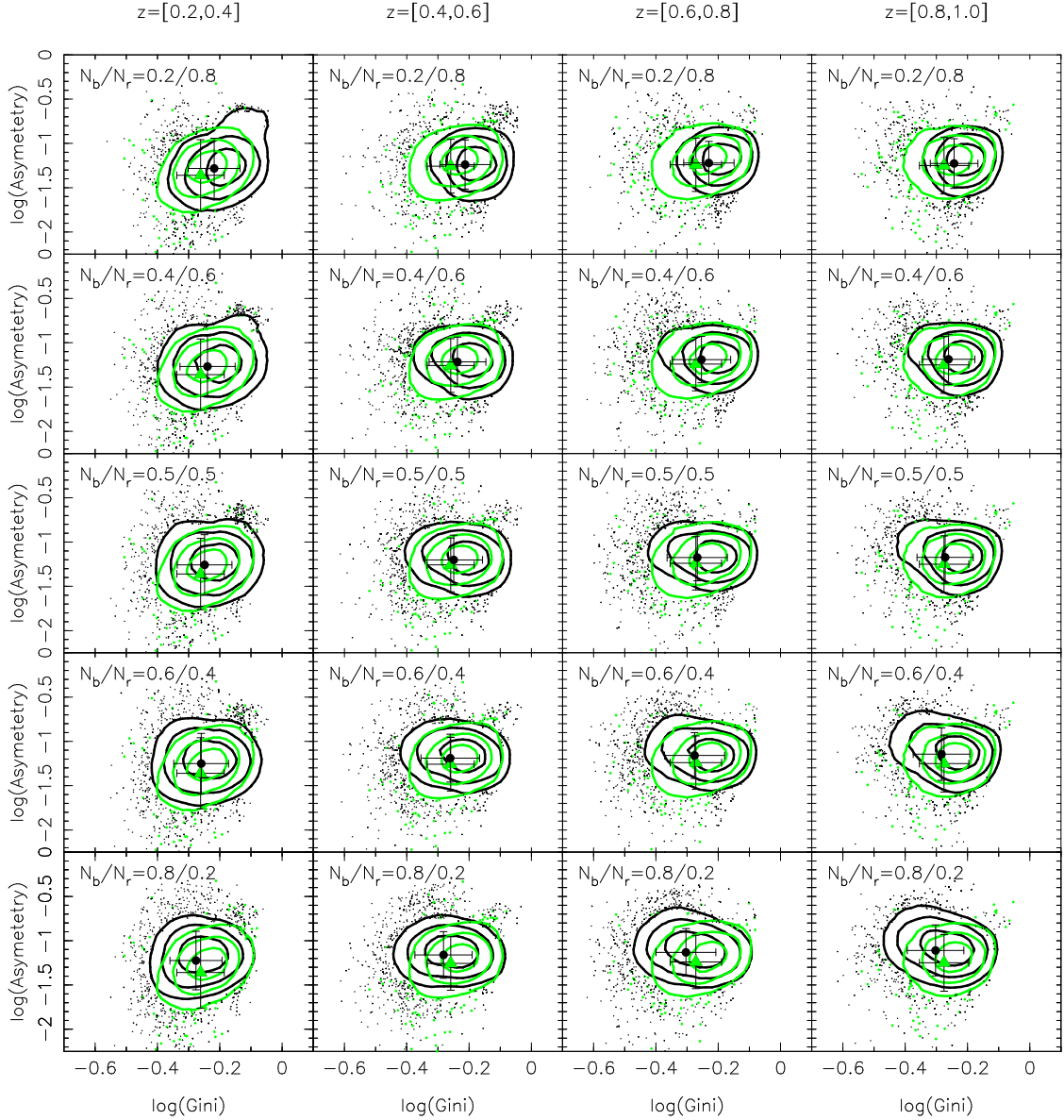}
\caption{ The Gini coefficient ($G$) vs. asymmetry ($A$) diagram for green
galaxies and mixed sample. Green galaxies are denoted as green dots. The green
and black contours enclose 30\%, 60\% and 80\% of the green and mixed galaxies,
respectively. Green triangle and black circle mark the average $Gini$
coefficient and $Asymmetry$ of green and mixed sample, respectively. The error
bars are the standard deviations of $Gini$ and $Asymmetry$ distribution for
each sample. The blue to red ratio $N_{b}/N_{r}$ is marked in the top left in
each panel.}
\end{figure*}

\begin{figure*}
\centering
\includegraphics[width=0.9\textwidth,angle=0]{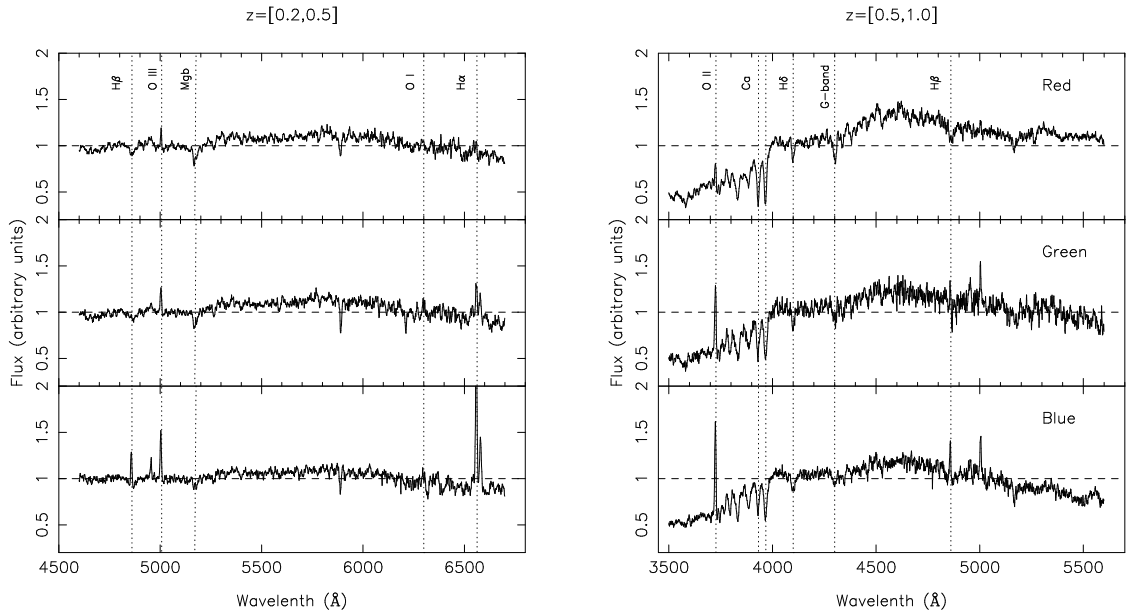}
\caption{ The stacked spectra of blue, green and red galaxies. The left panels
show the stacked spectra at the low redshift bin z=[0.2, 0.5]. The blue, green
and red spectrum are stacked by 378, 86, 390 spectra, respectively. The spectra
are normalized to the average flux between 5050 \AA\, and 5100\AA. The vertical
dotted lines show the ordinary emission lines or absorption lines. The right
panels show the stacked spectra at the high redshift bin z=[0.5, 1.0]. The red,
green and blue spectrum are stacked by 456,118 and 348 spectra, respectively,
and normalized to the average flux between 4050 \AA\, and 4100 \AA.}
\end{figure*}

\begin{figure}
\centering
\includegraphics[width=0.9\columnwidth, angle=0]{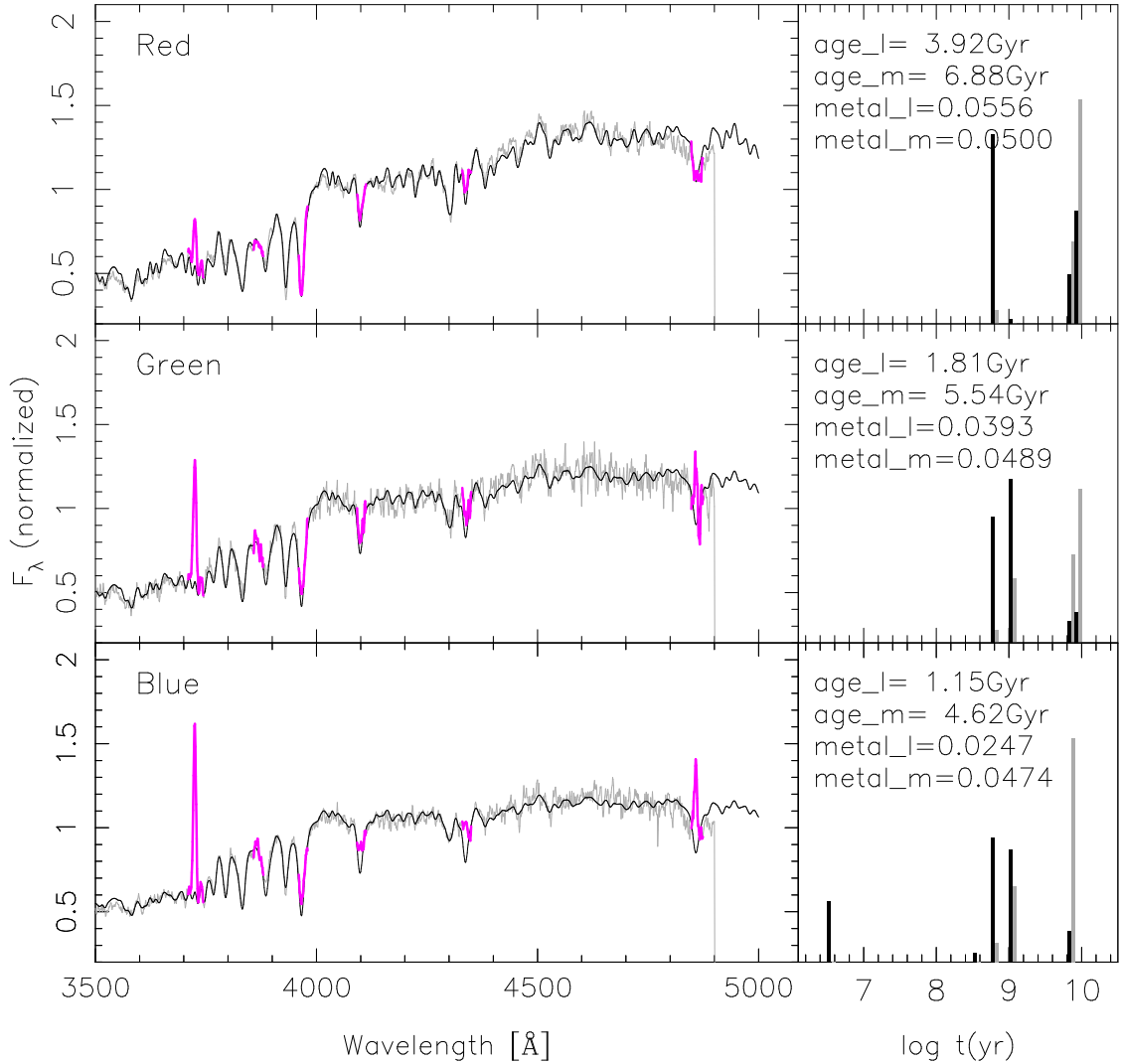}
\caption{
The results of STARLIGHT fitting on the stacked spectra. In the left
panels, the stacked spectra are shown in grey lines, and the best
fitting spectra are shown in black solid lines. The pink lines show
the masked regions.  In the right panels, the solid black histograms
show the flux-weighted SSP vectors, while the grey histograms show
the mass-weighted vectors. The flux-weighted age, mass-weighted age,
flux-weighted metallicity and mass-weighted metallicity are shown from
top to bottom.}
\end{figure}

\begin{figure*}
\centering
\includegraphics[width=0.9\textwidth,angle=0]{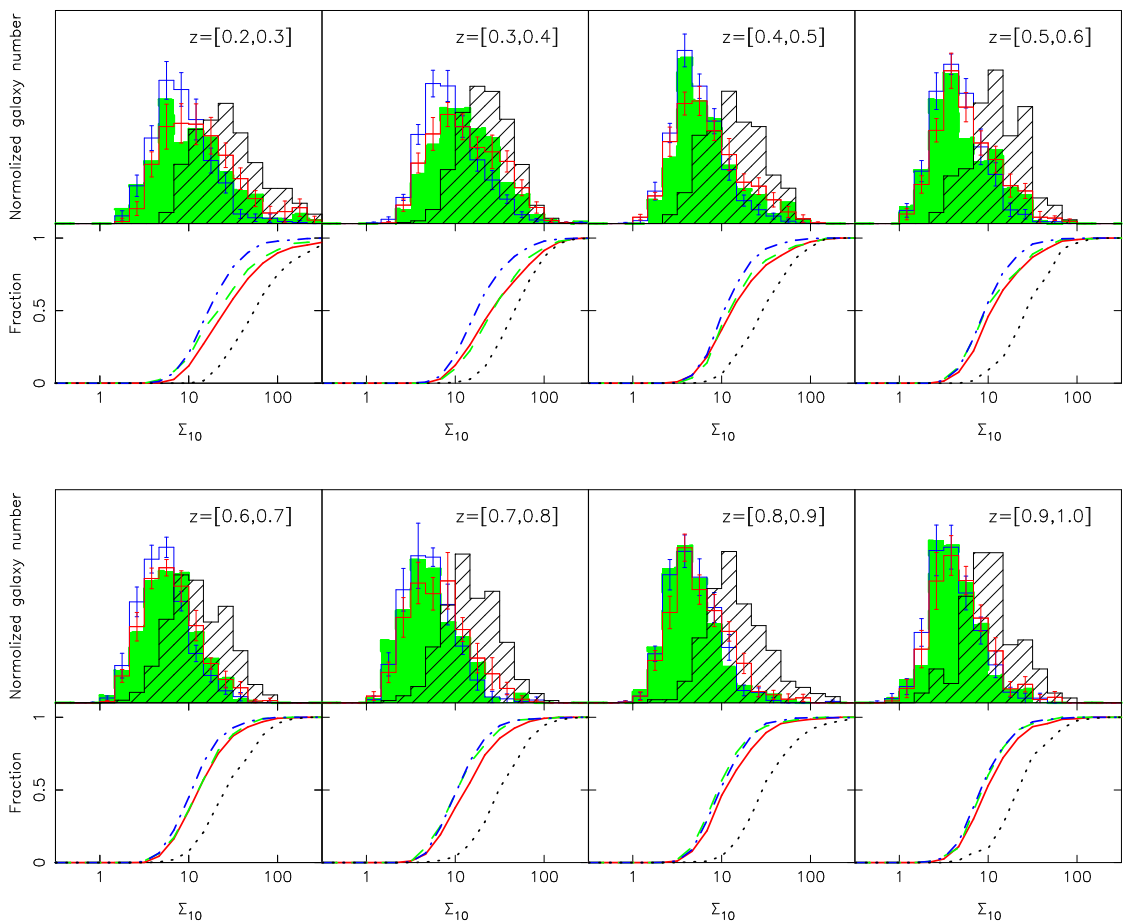}
\caption{
$\Sigma_{10}$ distributions of blue (blue solid), green (green filled),
red galaxies (red solid) and X-ray group members (black shaded). The
histogram of blue, red and group members have been normalized to have same
effective area with that of green galaxies. The blue and red histograms
are averaged over the $\Sigma_{10}$ distribution of 15 different comparison
samples, and errorbars are the standard deviations of galaxy number in
that $\Sigma_{10}$ bin. The smaller panels bellow each histogram show
the cumulative fraction distributions (blue galaxies: dot-dashed lines;
green galaxies: dashed lines; red galaxies: solid lines; X-ray group
members: dotted lines.).}
\end{figure*}

\begin{figure*}
\centering
\includegraphics[width=0.9\textwidth, angle=0]{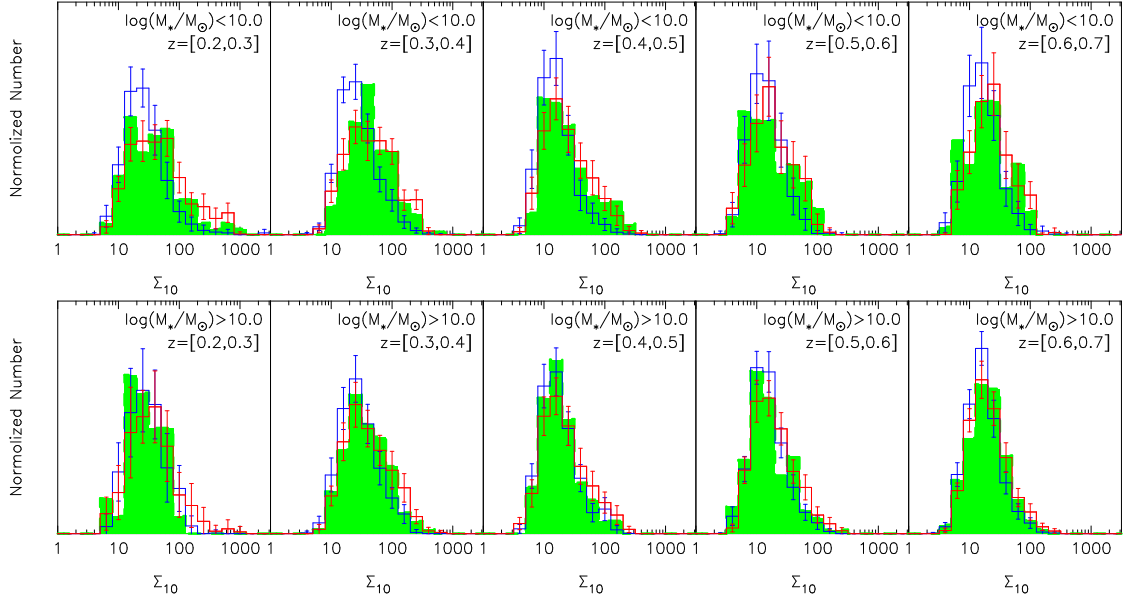}
\caption{
$\Sigma_{10}$ distributions of blue (blue dashed), green (green filled),
red galaxies (red solid) at $z=[0.2,0.7]$, shown into 2 mass bin:
log$(M_\ast/M_{\sun})$=[10.0,11.0] and log$(M_\ast/M_{\sun})$<10.0. }
\end{figure*}

\begin{figure*}
\centering
\includegraphics[width=0.9\textwidth, angle=0]{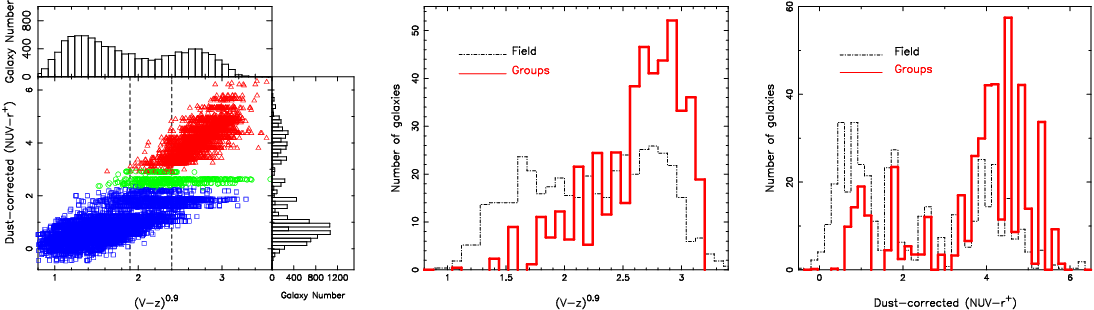}
\caption{
Left panel: the color-color diagram (k-corrected $(V-z)^{0.9}$ vs.
\nuvrcorr\ ) at $0.8<z<1.0$, along with the histograms of k-corrected
$(V-z)^{0.9}$ color and \nuvrcorr\ . Blue, green and red galaxies are
marked as blue squares, green circles and red triangles, respectively. The
dashed vertical lines indicate the green valley definition of
\citet{Balogh 2011}. It is clear that the two criteria are different and
only about 30\% green galaxies selected with \citet{Balogh 2011} criterion
meet our selection criterion.  middle panel: The k-corrected $(V-z)^{0.9}$
color histograms for group members ($P_{\rm mem}\geq 0.1$, red solid
line) and field galaxies ($P_{\rm mem}=0.0$ and log($\Sigma_{10})<0.5$,
black dot-dashed line) with $M_{\ast}>10^{10.1}M_{\odot}$. The field
galaxy histogram has been normalized and has the same area as the
group galaxy histogram. Right panel: \nuvr\ color distribution for
 field and group galaxies with $M_{\ast}>10^{10.1}M_{\odot}$.}
\end{figure*}


\begin{thebibliography}{99}
\bibitem[\protect\citeauthoryear{Abraham et al.}{1996}]{Abraham 1996}
Abraham, R. G., Tanvir N. R., et al., 1996, MNRAS, 279, L47

\bibitem[\protect\citeauthoryear{Aird et al.}{2012}]{Aird 2012}
Aird, J. et al., 2012, ApJ, 746, 90

\bibitem[\protect\citeauthoryear{Alonso-Herrero et al.}{2008}]{Alonso 2008}
Alonso-Herrero, A., et al., 2008, ApJ, 677, 127

\bibitem[\protect\citeauthoryear{Alonso et al.}{2012}]{Alonso 2012}
Alonso, S., Mesa, V, Padilla, N. \& Lambas, D. G., 2012, A\&A, 539, 46
\bibitem[\protect\citeauthoryear{Baldry et al.}{2004}]{Baldry 2004}Baldry, I.
K., et al., 2004, ApJ, 600, 681
\bibitem[\protect\citeauthoryear{Baldwin et al.}{1981}]{Baldwin 1981}Baldwin, J.
A, Philips, M. M. \& Terlevich R. 1981, PASP, 93, 5
\bibitem[\protect\citeauthoryear{Balogh et al.}{2011}]{Balogh 2011}Balogh, M. L
et al., 2011, MNRAS, 412, 2303
\bibitem[\protect\citeauthoryear{Bell et al.}{2004}]{Bell 2004}Bell, E. L, et
al., 2004, ApJ, 608, 752
\bibitem[\protect\citeauthoryear{Bell et al.}{2012}]{Bell 2012}Bell, E. L, et
al., 2012, ApJ, 753, 167

\bibitem[\protect\citeauthoryear{Bertoldi et al.}{2007}]{Bertoldi
2007}Bertoldi, F., Carilli, C., Aravena, M., et al., 2007, ApJS, 172, 132

\bibitem[\protect\citeauthoryear{Blanton et al.}{2003}]{Blanton 2003}Blanton M.
R., Hogg D. W., et al. 2003, ApJ, 594, 18
\bibitem[\protect\citeauthoryear{Brammer et al.}{2009}]{Brammer 2009}Brammer G.
B., Whitaker K. E., et al, 2009, ApJ, 706, L173
\bibitem[\protect\citeauthoryear{Brammer et al.}{2011}]{Brammer
2011}Brammer G. B, Whitaker K. E., et al, 2011, ApJ, 739, 24
\bibitem[\protect\citeauthoryear{Barazza et al.}{2002}]{Barazza
2002}Barazza, F. D, Binggeli, B. \& Jerjen H., 2002, A\&A, 391, 823
\bibitem[\protect\citeauthoryear{Bruzual \& Charlot}{2003}]{Bruzual 2003}Bruzual G.,
Charlot S., 2003, MNRAS, 344, 1000
\bibitem[\protect\citeauthoryear{Bolzonella et al.}{2010}]{Bolzonella 2010}Bolzonella, M. et al., 2010, A\&A, 524A, 76B
\bibitem[\protect\citeauthoryear{Bongiorno et al.}{2012}]{Bongiorno 2012}Bongiorno, A. et al., 2012, MNRAS,
427,3103
\bibitem[\protect\citeauthoryear{Boselli \& Gavazzi}{2006}]{Boselli
2006}Boselli, A. \& Gavazzi G., 2006, PASP, 118, 517
\bibitem[\protect\citeauthoryear{Bower et al.}{2006}]{Bower 2006}Bower, R. G et al., 2006,
MNRAS, 370, 645
\bibitem[\protect\citeauthoryear{Bundy et al.}{2008}]{Bundy
2008}Bundy, K. et al., 2008, ApJ, 681, 931
\bibitem[\protect\citeauthoryear{Bundy et al.}{2010}]{Bundy
2010}Bundy K. et al., 2010, ApJ, 719, 1969
\bibitem[\protect\citeauthoryear{Butcher \& Oemler}{1978}]{Butcher 1978}Butcher H.\& Oemler A. Jr., 1978, ApJ, 226, 559
\bibitem[\protect\citeauthoryear{Calzetti et al.}{2000}]{Calzetti 2000}Calzetti. D, et al., 2000, ApJ,
533, 682
\bibitem[\protect\citeauthoryear{Capak et al.}{2007a}]{Capak 2007a}Capak. P,
Abraham R. et al., 2007a, ApJS, 172, 99
\bibitem[\protect\citeauthoryear{Capak et al.}{2007b}]{Capak 2007b}Capak. P,
Abraham, R. et al., 2007b, ApJS, 172, 284

\bibitem[\protect\citeauthoryear{Cardelli, Clayton \& Mathis}{1989}]{Cardelli 1989}Cardelli, J. A., Clayton, G. C., Mathis, J, S., 1989, ApJ, 345, 245
\bibitem[\protect\citeauthoryear{Chen et al.}{2009}]{Chen
2009}Chen Y. M, et al., 2009, MNRAS, 393, 406C
\bibitem[\protect\citeauthoryear{Cheung et al.}{2012}]{Cheung
2012}Cheung, E, et al., 2012, ApJ, 760, 131C
\bibitem[\protect\citeauthoryear{Cid Fernandes et al.}{2005}]{Cid 2005}Cid Fernandes. R., Mateus, A., Sodr\'{e}, L., Stasi\'{n}ska G., Gomes J. M., 2005, MNRAS, 358, 363
\bibitem[\protect\citeauthoryear{Cluver et al.}{2013}]{Cluver
2013}Cluver M E, et al., 2013, ApJ, 765, 93
\bibitem[\protect\citeauthoryear{Coil et al.}{2008}]{Coil
2008}Coil, A. L., et al., 2008, ApJ, 672, 153
\bibitem[\protect\citeauthoryear{Coil et al.}{2009}]{Coil
2009}Coil, A. L., et al., 2009, ApJ, 701, 1484
\bibitem[\protect\citeauthoryear{Conselice et al.}{2000}]{Conselice
2000}Conselice, C. J., Bershady, M. A. \& Jangren, A. 2000, ApJ, 529, 886
\bibitem[\protect\citeauthoryear{Cooper et al.}{2006}]{Cooper
2006}Cooper, M. C. Newman J. A., et al., 2006, MNRAS, 370, 198
\bibitem[\protect\citeauthoryear{Dale \$ Helou}{2002}]{Dale 2002}Dale, D.A. \& Helou, G, 2002, ApJ,
576, 159
\bibitem[\protect\citeauthoryear{da Silva et al.}{2011}]{dasilva 2011}da Silva, R. L ., et al., 2011, ApJ, 735, 54D
\bibitem[\protect\citeauthoryear{Donley et al.}{2012}]{Donley 2012}Donley, J.L. et
al., 2012, ApJ, 748, 142
\bibitem[\protect\citeauthoryear{Dressler et al.}{1997}]{Dressler
1997}Dressler, A .et al, 1997, ApJ, 490, 577
\bibitem[\protect\citeauthoryear{Driver et al.}{2006}]{Driver 2006}Driver, S. P.,
et al., 2006, MNRAS, 368, 414

\bibitem[\protect\citeauthoryear{Eisenstein et al.}{2003}]{Eisenstein
2003}Eisenstein, D. J., Hogg, D. W., et al, 2003, ApJ, 585, 694E
\bibitem[\protect\citeauthoryear{Ellison et al.}{2010}]{Ellison
2010}Ellison, S. L., et al, 2010, MNRAS, 407, 1514
\bibitem[\protect\citeauthoryear{Fasano et al.}{2000}]{Fassano
2000}Fassano, G .et al, 2000, ApJ, 542, 673
\bibitem[\protect\citeauthoryear{Faber et al.}{2007}]{Faber 2007}Faber, S. M. et
al., 2007, ApJ, 665, 265
\bibitem[\protect\citeauthoryear{Fang et al.}{2012}]{Fang 2012}
Fang, J, J., Faber, S. M, Salim, S, et al., 2012, ApJ, 761, 23

\bibitem[\protect\citeauthoryear{Feruglio et al.}{2010}]{Feruglio 2010}
Feruglio, C., Maiolino, R., Piconcelli, E. et al, 2010, A\&A, 518, 155

\bibitem[\protect\citeauthoryear{George et al.}{2011}]{George 2011}
George, S. et al., 2011, ApJ, 742, 125

\bibitem[\protect\citeauthoryear{Giodini et al.}{2012}]{Giodini 2012}
Giodini, M. G. et al.,2012, A\&A, 538, 104
\bibitem[\protect\citeauthoryear{Glasser}{1962}]{Glasser 1962}
Glasser, G. J., 1962, J. Amer. Stat. Assoc., 57, 648

\bibitem[\protect\citeauthoryear{Goncalves et al.}{2012}]{Goncalves 2012}
Goncalves, T. S. et al., 2012, ApJ, 759, 67
\bibitem[\protect\citeauthoryear{Haines et al.}{2006}]{Haines
20086}Haines, C. P. et al., 2006, ApJ, 657L, 21H
\bibitem[\protect\citeauthoryear{Haines et al.}{2006}]{Haines
2006}Haines, C. P., et al. 2006, ApJ, 647, L21
\bibitem[\protect\citeauthoryear{Haines et al.}{2008}]{Haines
2008}Haines, C. P., Gargiulo A. \& Merluzzi, P., 2008, MNRAS, 385, 1201
\bibitem[\protect\citeauthoryear{Hasinger et al.}{2007}]{Hasinger
2007}Hasinger G., Cappelluti N., et al., 2007, ApJS, 172, 29
\bibitem[\protect\citeauthoryear{Ideue et al.}{2012}]{Ideue
2012}Ideue, Y., Taniguchi, Y. et al., 2012, ApJ, 747, 42
\bibitem[\protect\citeauthoryear{Ilbert et al.}{2009}]{Ilbert 2009}Ilbert O., Capak P.,
 Salvato M. et al., 2009, ApJ, 690, 1236
\bibitem[\protect\citeauthoryear{Ilbert et al.}{2010}]{Ilbert 2010}Ilbert,
 O., Salvato, M. et al., 2010 ApJ, 709, 644

\bibitem[\protect\citeauthoryear{Kauffmann et al.}{2003}]{Kauffmann
2003}Kauffmann, G, Heckman, T. M, White, S.D.M, et al., 2003, MNRAS, 341, 33
\bibitem[\protect\citeauthoryear{Kauffmann et al.}{2003b}]{Kauffmann
2003b}Kauffmann G, Heckman T. M, Tremonti C, et al., 2003, MNRAS, 346, 1055
\bibitem[\protect\citeauthoryear{Kauffmann et al.}{2012}]{Kauffmann
2012}Kauffmann, G, Cheng, Li, Jian, Fu, et al., 2012, MNRAS, 422, 997
\bibitem[\protect\citeauthoryear{Kawata \& Mulchaey}{2008}]{Kawata 2008}Kawata, D.\& Mulchaey, J. S., 2008, ApJ, 672, L103
\bibitem[\protect\citeauthoryear{Kenneth et al.}{2011}]{Kenneth 2011}Kenneth C. W., et al., 2011, ApJ, 728, 119
\bibitem[\protect\citeauthoryear{Koekemoer et al.}{2007}]{Koekemoer
2007}Koekemoer, A. M., Aussel, H. et al., 2007, ApJS, 172,196
\bibitem[\protect\citeauthoryear{Kong et al.}{2009}]{Kong
2009}Kong, X., Fang, G., Arimoto, N., \& Wang, M.\ 2009, ApJ, 702, 1458

\bibitem[\protect\citeauthoryear{Kov$\rm \breve{a}$c et al.}{2010}]{Kovac
2010}Kov$\rm \breve{a}$c., et al., 2010, ApJ, 718, 86
\bibitem[\protect\citeauthoryear{Lacy et al.}{2004}]{Lacy
2004}Lacy, M. et al., 2004, ApJS, 154, 166
\bibitem[\protect\citeauthoryear{Lambas et al.}{2003}]{Lambas
2003}Lambas, D. G., Tissera, P. B., Alonso M. S, Coldwell, G., 2003, MNRAS,
346, 1189
\bibitem[\protect\citeauthoryear{Li et al.}{2008}]{Li
2008}Li, C., Kauffmann, G., Heckman, T. M., et al., 2008, MNRAS, 385, 1903
\bibitem[\protect\citeauthoryear{Lilly et al.}{2007}]{Lilly 2007}Lilly, S. J., Le F\textmd{$\grave{e}$}vre, Renzini, A., et al., 2007, ApJS, 172, 70

\bibitem[\protect\citeauthoryear{Lotz et al.}{2004}]{Lotz
2004}Lotz, J. M., Primack, J. \& Madau, P., 2004, AJ, 128, 163
\bibitem[\protect\citeauthoryear{Lotz et al.}{2006}]{Lotz
2006}Lotz, J. M., Madau, P. et al., 2006, ApJ, 636, 592

\bibitem[\protect\citeauthoryear{Lubin et al.}{2002}]{Lubin
2002}Lubin, L. M., Oke J. B., \& Postman, M., 2002, AJ, 124, 1905L
\bibitem[\protect\citeauthoryear{Lusso et al.}{2011}]{Lusso 2011} Lusso, E., Comastri, A., Vignali, C., et al.\ 2011, A\&A, 534, A110
\bibitem[\protect\citeauthoryear{Marino et al.}{2011}]{Marino 2011}Marino, A.
Bianchi, L., Rampazzo, R., et al., 2011, ApJ, 736, 154
\bibitem[\protect\citeauthoryear{Martig et al.}{2009}]{Martig 2009}Martig, M.
Bournaud, F., Teyssier, R. \& Dekel, A., 2009, ApJ, 707, 250
\bibitem[\protect\citeauthoryear{Martin et al.}{2007}]{Martin 2007}Martin, D. C.
et al., 2007, ApJS, 173, 342
\bibitem[\protect\citeauthoryear{McCathy et al.}{2008}]{McCarthy 2008}McCarthy,
I. G. et al., 2008, MNRAS, 38,593
\bibitem[\protect\citeauthoryear{McNamara et al.}{2000}]{McNamara 2000}McNamara
B. R. et al., 2000, ApJ, 534, L135
\bibitem[\protect\citeauthoryear{McNamara et al.}{2007}]{McNamara 2007}McNamara, B. R. \& Nulsen, P. E. J., 2007, ARA\&A, 45,
117
\bibitem[\protect\citeauthoryear{Mendez et al.}{2011}]{Mendez 2011}Mendez, A.
J., et al. 2011, ApJ, 736, 110
\bibitem[\protect\citeauthoryear{Mendez et al.}{2013}]{Mendez 2013}Mendez, A.
J., et al. 2013, ApJ, 770, 40
\bibitem[\protect\citeauthoryear{Mobasher et al.}{2007}]{Mobasher
2007}Mobasher, B. et al., 2007, ApJS, 172, 117
\bibitem[\protect\citeauthoryear{Moran et al.}{2007}]{Moran 2007}Moran, S. M.
Ellis, R. S. et al., 2007, ApJ, 671, 1503
\bibitem[\protect\citeauthoryear{Moresco et al.}{2010}]{Moresco
2010}Moresco, M., Pozzetti, L. et al., 2010, A\&A, 524, 67
\bibitem[\protect\citeauthoryear{Moustakas et al.}{2013}]{Moustakas
2013}Moustakas, J., Coil, A. L., Aird, J. et al., 2013, ApJ, 767, 50
\bibitem[\protect\citeauthoryear{Nandra et al.}{2007}]{Nandra
2007}Nandra, K., Georgakakis, A. et al., 2007, ApJ, 660, L11
\bibitem[\protect\citeauthoryear{Nikolic et al.}{2004}]{Nikolic
2004}Nikolic, B, Cullen, H \& Alexander, P., 2004, MNRAS, 355, 874
\bibitem[\protect\citeauthoryear{Page et al.}{2012}]{Page 2012}Page, M. J.,
Symeonidis, M, Vieira, J. D., et al., 2012, Nature, 485, 213
\bibitem[\protect\citeauthoryear{Pan et al.}{2012}]{Pan 2012}Pan, Z. Z., Yuan, Q.
R., Kong, X. et al., 2012, MNRAS, 421, 36
\bibitem[\protect\citeauthoryear{Patton et al.}{2011}]{Patton 2011}Patton., et al., 2011, MNRAS, 412, 591
\bibitem[\protect\citeauthoryear{Peng et al.}{2010}]{Peng 2010}Peng Y. J., et al., 2010, ApJ, 721, 193
\bibitem[\protect\citeauthoryear{Perez et al.}{2009}]{Perez 2009}Perez, J. et
al., 2009, MNRAS, 399, 1157
\bibitem[\protect\citeauthoryear{Polletta et al.}{2007}]{Polletta 2007}Polletta, M.
 et al., 2007 ApJ, 663, 81
\bibitem[\protect\citeauthoryear{Pozzetti et al.}{2010}]{Pozzetti
 2010}Pozzetti, L. et al., 2010, A\&A, 523, A13
\bibitem[\protect\citeauthoryear{Prevot}{1984}]{Prevot
 1984}Prevot, M, L., et al, A\&A, 132, 389
\bibitem[\protect\citeauthoryear{Rampazzo et al.}{2007}]{Rampazzo
2007}Rampazzo, R., Marino, A., Tantalo, R., et al., 2007, MNRAS, 381, 245
\bibitem[\protect\citeauthoryear{Robertson et al.}{2006}]{Robertson
2006}Robertson, B. et al., 2006, ApJ, 645, 986
\bibitem[\protect\citeauthoryear{Rosario et al.}{2013a}]{Rosario
2013a}Rosario, D. J et al., 2013, ApJ, 763, 59
\bibitem[\protect\citeauthoryear{Rosario et al.}{2013b}]{Rosario
2013b}Rosario, D. J et al., 2013, ApJ accepted, arXiv:1302.1202v2
\bibitem[\protect\citeauthoryear{Salim et al.}{2005}]{Salim
2005}Salim, S. et al., 2005, ApJ, 619, L39
\bibitem[\protect\citeauthoryear{Salim et al.}{2007}]{Salim
2007}Salim, S., et al., 2007, ApJS, 173, 267
\bibitem[\protect\citeauthoryear{Salim et al.}{2009}]{Salim 2009}Salim, S. et
al., 2009, ApJ, 700, 161
\bibitem[\protect\citeauthoryear{Salim et al.}{2012}]{Salim 2012}Salim, S. Fang, J. J, Rich, R. M, et al., 2012, ApJ, 700, 161
\bibitem[\protect\citeauthoryear{Sanders et al.}{2007}]{Sanders
2007}Sanders, D. B., Salvato, M, Aussel H., et al., 2007, ApJS, 172, 86
\bibitem[\protect\citeauthoryear{Salvato et al.}{2009}]{Salvato
2009}Salvato, M. et al., 2009, ApJ, 690, 1250
\bibitem[\protect\citeauthoryear{Scarlata et al.}{2007}]{Scarlata 2007}Scalata, C., Carollo, C. et al. 2007, ApJS, 172, 406
\bibitem[\protect\citeauthoryear{Scoville et al.}{2007}]{Scoville
2007}Scoville, N., Aussel, H., Brusa, M., et al., ApJS, 172, 1

\bibitem[\protect\citeauthoryear{Shiavon}{2007}]{Shiavon 2007}Schiavon R. P., 2007, ApJS, 171, 146
\bibitem[\protect\citeauthoryear{Shu et al.}{2012}]{Shu
2012}Shu, Y. P, et al., AJ, 143, 90S
\bibitem[\protect\citeauthoryear{Silverman et al.}{2009}]{Silverman 2009} Silverman, J.~D.,
Lamareille, F., Maier, C., et al.\ 2009, ApJ, 696, 396
\bibitem[\protect\citeauthoryear{Smith et al.}{2010}]{Smith 2010}Smith, R. J., et al., 2010, MNRAS, 408, 1417
\bibitem[\protect\citeauthoryear{Snyder et al.}{2011}]{Snyder 2011}Snyder, G. F ., et al., 2011, ApJ, 741, 77s
\bibitem[\protect\citeauthoryear{Springel et al.}{2005}]{Springel
2005}Springel, V., Di, Matteo T., \& Hernquist, L., 2005, ApJ, 620, L79
\bibitem[\protect\citeauthoryear{Stern et al.}{2005}]{Stern
2005}Stern, D., et al., 2005, ApJ, 631, 163
\bibitem[\protect\citeauthoryear{Strateva et al.}{2001}]{Strateva
2001}Strateva, I. V et al., 2001, AJ, 122, 1861
\bibitem[\protect\citeauthoryear{Sturm et al.}{2011}]{Sturm
2011}Sturm, E. et al., 2011, ApJ, 733, 16

\bibitem[\protect\citeauthoryear{Tasca et al.}{2009}]{Tasca
2009}Tasca, L. A. M, et al., 2009, A\&A, 503, 379
\bibitem[\protect\citeauthoryear{Thomas et al.}{2005}]{Thomas 2005}Thomas, D., Maraston, C., Bender, R., Mendes de Oliveira C., 2005, ApJ, 621, 673
\bibitem[\protect\citeauthoryear{Thomas et al.}{2010}]{Thomas 2010}Thomas, D., Maraston, C., Schawinski, K., et al., 2010, MNRAS, 404,
1775
\bibitem[\protect\citeauthoryear{Tremonti et al.}{2007}]{Tremonti 2007}Tremonti, C. A,
Moustakas, J, Diamond, S., \& Aleksandar, M., 2007, ApJ, 663, 77
\bibitem[\protect\citeauthoryear{Thilker et al.}{2010}]{Thilker 2010}Thilker,
D., et al., 2010, ApJ, 714, L171
\bibitem[\protect\citeauthoryear{van den Bosch et al.}{2008}]{vandenBosch
2008}van den Bosch, F. C., Aquino, D., Yang, X. H. et al., 2008, MNRAS, 387, 79
\bibitem[\protect\citeauthoryear{van der Wel et al.}{2010}]{vanderwel 2010}van der Wel, A., et al., 2010, ApJ, 714, 1779
\bibitem[\protect\citeauthoryear{Vogt et al.}{2004}]{Vogt 2004}Vogt, N. P,
Haynes, M. P, Giovanelli, R. \& Herter, T., 2004, AJ, 127, 3300

\bibitem[\protect\citeauthoryear{Wang et al.}{2012}]{Wang
2012}Wang, T. et al., 2012, ApJ, 752, 134
\bibitem[\protect\citeauthoryear{Weinmann et al.}{2010}]{Weinmann 2010}Weinmann, S. M., et al., 2010, MNRAS, 406,
2249
\bibitem[\protect\citeauthoryear{Whitaker et al.}{2010}]{Whitaker
2010}Whitaker, K. E., et al., 2010., ApJ, 719, 1715
\bibitem[\protect\citeauthoryear{Willmer et al.}{2006}]{Willmer 2006}Willmer
C. N. A, Faber, S. M. et al., 2006, ApJ, 647, 853
\bibitem[\protect\citeauthoryear{Williams et al.}{2009}]{Williams 2009}Williams
R, J., et al., 2009, ApJ, 691, 1879
\bibitem[\protect\citeauthoryear{Wyder et al.}{2007}]{Wyder 2007}Williams
R, J., et al., 2007, ApJS, 173, 293
\bibitem[\protect\citeauthoryear{Xue et al.}{2010}]{Xue 2010}Xue, Y. Q., Brandt, W. N. et al., 2010, ApJ, 720, 368
\bibitem[\protect\citeauthoryear{Yan et al.}{2011}]{Yan
2011}Yan, R. et al., 2011, ApJ, 728, 38
\bibitem[\protect\citeauthoryear{Zamojski et al.}{2007}]{Zamojski
2007}Zamojski,M. A., Schiminovich, D. et al., 2007, ApJS, 172, 468 6
\bibitem[\protect\citeauthoryear{Zehavi et al.}{2011}]{Zehavi 2011}Zehavi
I., Zheng Z. et al., 2011, ApJ, 736, 59
\bibitem[\protect\citeauthoryear{Zhu et al.}{2010}]{Zhu 2010}Zhu, G. T., Blanton, M. R. \& Moustakas J., 2010, ApJ, 722, 491

\end{thebibliography}
\end{document}